\documentclass[11pt,reqno]{amsproc}
\usepackage[colorinlistoftodos,shadow]{todonotes}
\usepackage[T1]{fontenc}
\usepackage{amsmath,amscd,amssymb}
\usepackage{cite}
\usepackage{color}
\usepackage{graphicx}
\usepackage{psfrag,epsfig}
\usepackage{multirow}
\usepackage{rotating}
\usepackage{fullpage}
\usepackage{subfigure}
\usepackage{enumerate}
\usepackage{comment}
\usepackage{lineno}
\usepackage{algorithmic}
\usepackage{algorithm}
\usepackage[small]{caption}
\usepackage[debug=false, colorlinks=true, pdfstartview=FitV, linkcolor=blue, citecolor=blue, urlcolor=blue]{hyperref}
\newtheorem{theorem}{Theorem}[section]

\newtheorem{remark}[theorem]{Remark}

\numberwithin{equation}{section}
\usepackage{chemarr}
\usepackage[version=3]{mhchem}
\newcommand*\patchAmsMathEnvironmentForLineno[1]{%
  \expandafter\let\csname old#1\expandafter\endcsname\csname #1\endcsname
  \expandafter\let\csname oldend#1\expandafter\endcsname\csname end#1\endcsname
  \renewenvironment{#1}%
     {\linenomath\csname old#1\endcsname}%
     {\csname oldend#1\endcsname\endlinenomath}}%
\newcommand*\patchBothAmsMathEnvironmentsForLineno[1]{%
  \patchAmsMathEnvironmentForLineno{#1}%
  \patchAmsMathEnvironmentForLineno{#1*}}%
\AtBeginDocument{%
\patchBothAmsMathEnvironmentsForLineno{equation}%
\patchBothAmsMathEnvironmentsForLineno{align}%
\patchBothAmsMathEnvironmentsForLineno{flalign}%
\patchBothAmsMathEnvironmentsForLineno{alignat}%
\patchBothAmsMathEnvironmentsForLineno{gather}%
\patchBothAmsMathEnvironmentsForLineno{multline}%
}

\title{Regression-based reduced-order models to predict transient thermal 
output for enhanced geothermal systems}
\author{M.~K.~Mudunuru$^{*}$, S.~Karra, D.~R.~Harp, G.~D.~Guthrie, 
and H.~S.~Viswanathan \\
{\scriptsize Earth and Environmental Sciences Division, 
Los Alamos National Laboratory, Los Alamos, NM 87545.} \\
}
\address{\small $^*$Corresponding author: Dr.~Maruti Kumar Mudunuru, 
Computational Earth Science Group (EES-16), Earth and Environmental Sciences 
Division, Los Alamos National Laboratory, Los Alamos, NM 87545. \\
\textbf{E-mail address:} maruti@lanl.gov \newline \newline
}
\date{\today}
\begin{document}
\maketitle
%
%
\section*{Abstract} 
Reduced-order modeling is a promising approach, as many phenomena 
can be described by a few parameters/mechanisms. 
An advantage and attractive aspect of a reduced-order 
model is that it is computational inexpensive to evaluate when compared 
to running a high-fidelity numerical simulation. A reduced-order 
model takes couple of seconds to run on a laptop while a high-fidelity 
simulation may take couple of hours to run on a high-performance computing 
cluster. The goal of this paper is to assess the utility of regression-based 
Reduced-Order Models (ROMs) developed from high-fidelity numerical 
simulations for predicting transient thermal power output for an enhanced 
geothermal reservoir while explicitly accounting for uncertainties in the 
subsurface system and site-specific details. Numerical simulations are 
performed based on equally spaced values in the specified range of model 
parameters. Key sensitive parameters are then identified from these simulations, 
which are fracture zone permeability, well/skin factor, bottom hole 
pressure, and injection flow rate. We found the fracture zone permeability 
to be the most sensitive parameter. The fracture zone permeability along 
with time, are used to build regression-based ROMs for the 
thermal power output. The ROMs are trained and validated using detailed 
physics-based numerical simulations. Finally, predictions from the ROMs 
are then compared with field data. We propose three different ROMs with 
different levels of model parsimony, each describing key and essential 
features of the power production curves. The coefficients 
in proposed regression-based ROMs are developed by minimizing a non-linear 
least-squares misfit function using Levenberg-Marquardt algorithm. The 
misfit function is based on the difference between numerical simulation 
data and reduced-order model.
ROM-1 is constructed based on polynomials upto fourth order. 
ROM-1 is able to accurately reproduce the power output of numerical simulations 
for low values of permeabilities and certain features of the field-scale 
data. ROM-2 is a model with more analytical functions consisting of polynomials
upto order eight, exponential functions and smooth approximations of Heaviside 
functions, and accurately describes the field-data. At higher permeabilities, 
ROM-2 reproduces numerical results better than ROM-1, however, there is a 
considerable deviation from numerical results at low fracture zone permeabilities.
ROM-3 consists of polynomials upto order ten, and is developed by taking the 
best aspects of ROM-1 and ROM-2. ROM-1 is relatively parsimonious than ROM-2 
and ROM-3, while ROM-2 overfits the data. ROM-3 on the other hand, provides 
a middle ground for model parsimony. Based on $\mathrm{R}^2$-values for training, 
validation, and prediction data sets we found that ROM-3 is better model than 
ROM-2 and ROM-1. For predicting thermal drawdown in EGS applications, 
where high fracture zone permeabilities (typically greater than $10^{-15} \, 
\mathrm{m}^{2}$) are desired, ROM-2 and ROM-3 outperform ROM-1. As per computational 
time, all the ROMs are $10^4$ times faster when compared to running a high-fidelity 
numerical simulation. This makes the proposed regression-based ROMs attractive for 
real-time EGS applications because they are fast and provide reasonably good predictions 
for thermal power output.
\\
\\
\textbf{Keywords:} Enhanced Geothermal Systems (EGS), Reduced-Order Models (ROMs),
thermal drawdown,  regression.
%

\section{INTRODUCTION AND PROBLEM DESCRIPTION}
\label{Sec:S1_Geothermal_Intro}
Enhanced Geothermal Systems (EGS) present a significant and 
long-term opportunity for widespread power production from 
new geothermal sources. EGS makes it possible to tap otherwise 
inaccessible thermal resources in areas that lack traditional 
geothermal systems. It is estimated that within the USA alone 
the electricity production potential of EGS is in excess of 
100GW. Hence, the efforts to model and predict the 
performance of EGS reservoirs under various reservoir conditions 
(such as formation permeability, reservoir temperature, existing 
fracture/fault connectivity, and in-situ stress distribution) are 
vital. In this paper, we present a study based on reduced-order 
modeling using data from the historic Fenton Hill Hot Dry Rock 
(HDR) project \cite{Brown_etal}. From a practical point of view, 
we are interested in developing fast models to examine the potential 
for supplying thermal energy at sustained rates for commercial operations.

Most of the existing studies related to EGS \cite{1984_Robinson_Tester_JGR_v89_p10374_p10384,
1988_Barton_Zoback_GRL_v15_p467_p470, 1989_Grisby_etal_Geothermics_v18_p629_p656,
1989_Grisby_etal_Geothermics_v18_p657_p676,1993_Rodrigues_etal_SGW,1994_Kruger_etal_SGW,
1995_Swenson_etal_WGC, 2016_Mudunuru_etal_arxiv_JI,1996_Roff_etal_IJRMMSGA_v25_p627_p639,2011_Fu_etal_SGW,2012_Ghassemi_GGE_v30_p647_p664,2012_Kelkar_etal_SGW,2014_McClure_Horne_IJRMMS_v72_p242_p260,2014_Norbeck_etal_SGW,
2014_Pandey_etal_Geothermics_v51_p46_p62,2015_Pandey_etal_Geothermics_v51_p46_p62,
2016_Guo_etal_Geothermics_v61_p46_p62} are based on high-fidelity numerical simulations. 
These simulations are performed to gain detailed understanding of the physical processes 
taking place in EGS reservoirs. However, such detailed numerical simulations are computationally 
expensive as they take hours to run on hundreds of processors, making them prohibitive 
for real-time applications. Herein, we shall take a different route to model and understand 
EGS systems based on reduced-order modeling. Reduced-order models are similar to analytical 
solutions. A major advantage of ROMs are that they take couple of seconds to run on a laptop 
as compared to high-fidelity numerical simulations. This makes them attractive for real-time 
and commercial applications. But the procedure to construct ROMs is different compared to 
constructing analytical solutions. In literature, they are various ways to develop reduced-order 
models \cite{2009_Cardoso_etal_IJNME_v77_p1322_p1350,2010_Cardoso_Durlofsky_JCPE_v229_p681_p700,
2011_He_etal_JCP_v230_p8313_p8341,2013_Pau_etal_CS_v17_p705_p721,2013_Pasetto_etal_WRR_v49_p3215_p3228,
2016_Mudunuru_SGW}. In our case, reduced-order models are developed and trained based on 
high-fidelity numerical simulations. ROMs developed from high-fidelity numerical simulation 
data are not actual reduction of the physical system but are proxy/surrogate models for 
certain quantities of interest such as thermal power production. As a result, there are 
no governing equations for such reduced-order models. In the following subsections, we 
briefly describe the field experiment and need for reduced-order models. 

\subsection{A brief description of the problem and field experiment}
\label{SubSec:S1_GTOCCS_ChallengeProblem}
The aim of this paper is to predict the thermal power output during the 
Long-Term Flow Test (LTFT) experiment of Phase II reservoir at the Fenton 
Hill HDR test site, located near Los Alamos, New Mexico \cite{2015_MarkWhite_etal_SGW,
2016_MarkWhite_etal_SGW,2016_SigneWhite_etal_SGW_GTO,2016_SigneWhite_etal_SGW_GTO1}. 
This Phase II reservoir was designed to test the HDR concept at temperatures 
and thermal production rates near those required for a commercial electrical 
power plant \cite{HDR_FentonHill_2015,Brown_etal}. One of the objectives of 
the current study is to include site-specific conditions in developing models.

The operation of LTFT experiment at Fenton Hill lasted for 39 months with 11 
months of active circulation through the reservoir. The Phase II reservoir 
comprised of a single injection well and single production well. The injection 
pressure was in the range of 25 MPa to 30 MPa and production backpressure ranged from 
8 MPa to 13 MPa. The injection and production mass flow rates ranged from 7.5 $\mathrm{kg}
\, \mathrm{s}^{-1}$ to 8.5 $\mathrm{kg} \, \mathrm{s}^{-1}$ and 5.5 $\mathrm{kg} \, 
\mathrm{s}^{-1}$ to 7.0 $\mathrm{kg} \, \mathrm{s}^{-1}$. The injection and production 
temperatures at the well-head ranged from 293 K ($20^{\mathrm{o}} \,\mathrm{C}$) to 
303 K ($30^{\mathrm{o}} \, \mathrm{C}$) and 438 K ($165^{\mathrm{o}} \, \mathrm{C}$) 
to 458 K ($185^{\mathrm{o}} \, \mathrm{C}$). Correspondingly, the bottomhole temperatures 
at the production well accounting for friction loss are between 453 K ($180^{\mathrm{o}}
\, \mathrm{C}$) and 498 K ($225^{\mathrm{o}} \, \mathrm{C}$). For more details on these data 
sets and other aspects (such as tracer data, microseismic data, fracture networks, fracture 
connectivity, injection and production wells entry points, and reactive-transport data), 
see References \cite{HDR_FentonHill_2015,Brown_etal}.

\subsection{Need for reduced-order models}
\label{SubSec:S1_GTOCCS_ROMs}
In recent years, model reduction techniques have proven to be powerful 
tools for solving various problems in geosciences. In reservoir management 
and decision-making, ROMs are considered efficient yet powerful techniques 
to address computational challenges associated with managing realistic reservoirs 
\cite{2015_Alghareeb_MIT}. Examples of some popular research and scientific 
endeavors on reduced-order modeling within the context of subsurface processes 
include Cardoso et al. \cite{2009_Cardoso_etal_IJNME_v77_p1322_p1350,
2010_Cardoso_Durlofsky_JCPE_v229_p681_p700}, He et al. \cite{2011_He_etal_JCP_v230_p8313_p8341}, 
Pau et al. \cite{2013_Pau_etal_CS_v17_p705_p721}, and Pasetto et al. 
\cite{2013_Pasetto_etal_WRR_v49_p3215_p3228}. Loosely speaking, the 
problem of model reduction is to replace a detailed physics-based model 
of a complex system (or a set of processes) by a much ``simpler'' and 
more computationally efficient model than the original model while still 
accurately predicting those aspects of the system that are of interest. 
There are several reasons ROMs are useful in the context of EGS, including:
\begin{itemize}
  \item [$\blacktriangleright$] ROMs facilitate in developing site-specific
    models which are fast to compute. In general, they can be also used as a 
    substitute model for parameter estimation instead of inverse modeling, 
    which may require repeated evaluation of forward models.
  \item [$\blacktriangleright$] ROMs can be used as numerical surrogates 
    to perform detailed sensitivity analysis and parametric studies, thereby 
    reducing the overall computational burden of high-fidelity numerical 
    simulations.
  \item [$\blacktriangleright$] In many-query applications such as EGS, a simple, 
    efficient and predictive model is required for field use. Such a model can 
    greatly reduce (or minimize) the associated operational costs, thereby maximizing 
    the EGS power output potential.
\end{itemize}

The ROMs developed in this paper are based on regression. They are a 
combination of polynomials of different degrees, sinusoidal function, exponential 
function, and logistic function. The coefficients of these polynomials and functions 
are obtained by training the ROMs on high-fidelity numerical simulation data.

\subsection{Objectives and outline of the paper}
\label{SubSec:S1_Geothermal_Scope}

Among the various potential outputs of interest from a full-physics based simulation 
for the challenge problem is the thermal power produced.
The objective of the study is to develop ROMs for an EGS 
to predict thermal power output. We are interested 
in developing a fast and  predictive ROM for thermal power 
output that accurately reproduces detailed 
simulations. To achieve this, the overall approach involves obtaining thermal 
power data from detailed physics-based 3D numerical simulations using the parallel subsurface 
flow simulator PFLOTRAN \cite{2015_Lichtner_etal_PFLOTRAN}. Approximate range of model 
input parameters are constructed based an educated guess of the fractured EGS system 
(see Kelkar et al. \cite[Section-3]{HDR_FentonHill_2015}). We use equally spaced 
values of the input parameters to generate power data for several cases.
The ROMs are constructed, trained and validated against PFLOTRAN numerical simulations.
These ROMs are then used to predict thermal power output and compared with a field thermal power data set
from LTFT Fenton Hill HDR Phase II experiment.

The paper is organized as follows: Section \ref{Sec:S2_Geothermal_ConceptModel} 
describes the physics-based conceptual model, which approximately models 
the Fenton Hill HDR Phase II reservoir. It also provides a brief overview 
of the governing equations for fluid flow, thermal drawdown, and numerical 
methodology to solve the coupled conservation equations. Assumptions in 
modeling these systems are also outlined. In Section \ref{Sec:S2_Geothermal_ConceptModel}, 
we also perform calibration of the material parameters using the field thermal
output data. These values serve as base case for the parameters.
Using a range of values around these base case values, sensitiviy analysis is 
performed using PFLOTRAN and is shown in  Section \ref{Sec:S2_Geothermal_ConceptModel}. 
Workflow for ROM development is also 
described in this section as well. Section \ref{Sec:S3_Geothermal_GE_NumModel} 
details a procedure to construct ROMs for thermal power 
output. Details of training and validation are provided in this section as well.
Predictive capabilities of ROMs with respect to thermal power output and comparison against 
field-data is also discussed. Finally, 
conclusions are drawn in Section \ref{Sec:S4_Geothermal_Conclusions}.

%

\section{CONCEPTUAL MODEL, GOVERNING EQUATIONS, AND NUMERICAL METHODOLOGY}
\label{Sec:S2_Geothermal_ConceptModel}
In this section, we briefly describe the detailed numerical simulations 
we used in the development of the ROMs, including those governing equations 
needed to model physical processes involved in heat extraction within a 
jointed reservoir. We then present a physics-based conceptual model for 
an EGS reservoir and corresponding boundary conditions. Finally, we describe 
a numerical methodology to solve the system of coupled partial differential 
equations using the subsurface simulator PFLOTRAN. PFLOTRAN solves a system 
of nonlinear partial differential equations describing multiphase, multicomponent, 
and multiscale reactive flow and transport in porous materials using 
finite volume method. Here, we shall restrict to solving the governing 
equations resulting for single phase fluid (water) flow and heat transfer 
processes. 

\subsection{Governing equations:~Fluid flow and heat transfer}
\label{SubSec:S2_GE}
In order to predict the heat extraction process, we solve the following 
set of governing equations. This includes balance of mass and balance of 
energy for fluid flow and thermal drawdown. The governing mass conservation 
equation for single phase saturated flow is given by:
\begin{align}
  \label{Eqn:Richards_Eqn}
  \frac{\partial \varphi \rho}{\partial t} + \mathrm{div}
  [\rho \mathbf{q}] = Q_w
\end{align}
where $\varphi$ is the porosity, $\rho$ is the fluid density $[\mathrm{kmol} \, 
\mathrm{m}^{-3}]$, $\mathbf{q}$ is the Darcy's flux $[\mathrm{m} \, \mathrm{s}
^{-1}]$, and $Q_w$ is the volumetric source/sink term $[\mathrm{kmol} \, \mathrm{m}
^{-3} \, \mathrm{s}^{-1}]$. The Darcy's flux is given as follows:
\begin{align}
  \label{Eqn:Darcy_Flux_Eqn}
  \mathbf{q} = - \frac{k}{\mu} \mathrm{grad}
  [P - \rho g z]
\end{align}
where $k$ is the intrinsic permeability $[\mathrm{m}^{2}]$, $\mu$ is 
the dynamic viscosity $[\mathrm{Pa} \, \mathrm{s}]$, $P$ is the pressure 
$[\mathrm{Pa}]$, $g$ is the gravity $[\mathrm{m} \, \mathrm{s}^{-2}]$, 
and $z$ is the vertical component of the position vector $[\mathrm{m}]$. 
The source/sink term is given as follows:
\begin{align}
  \label{Eqn:SourceSink_Richards_Eqn}
  Q_w = \frac{q_M}{W_w} \delta \left(\mathbf{x} - 
  \mathbf{x}_{{\text{\tiny {ss}}}} \right), 
  \quad \mathrm{where} \; \; 
  q_M = \frac{\Gamma_{\text{\tiny {well}}} \rho}{\mu} 
  \left(P - P_{\text{\tiny {bhp}}} \right)
\end{align}
where $q_M$ is the mass flow rate $[\mathrm{kg} \, \mathrm{m}^{-3} 
\, \mathrm{s}^{-1}]$, $W_w$ is the formula weight of water $[\mathrm{kg} 
\, \mathrm{kmol}^{-1}]$, $\Gamma_{\text{\tiny {well}}}$ denotes the skin/well factor 
(which regulates the mass flow rate in the production well) [unit-less], 
$\mathbf{x}_{{\text{\tiny {ss}}}}$ denotes the location of the source/sink, 
$P_{\text{\tiny {bhp}}}$ is the bottom hole pressure of the production 
well, and $\delta(\bullet)$ denotes the Dirac delta distribution 
\cite{Evans_PDE}. The governing equation for energy conservation 
to model thermal drawdown and corresponding heat extraction processes 
is given as follows:
\begin{align}
  \label{Eqn:BOE_Eqn}
  \frac{\partial}{\partial t} \left(\varphi \rho U 
  + \left(1 - \varphi \right) \rho_{\text{\tiny {rock}}} 
  c_{\text{\tiny {p,rock}}} T \right) + \mathrm{div}
  [\rho \mathbf{q}H - \kappa \mathrm{grad}[T]] = Q_e
\end{align}
where $U$ is the internal energy of the fluid, $\rho_{\text{\tiny 
{rock}}}$ is the true density of the rock (or rock grain density), 
$c_{\text{\tiny {p,rock}}}$ is the true heat capacity of the rock 
(or rock grain heat capacity), $T$ is the temperature of the fluid, 
$H$ is the enthalpy of the fluid, $\kappa$ is the thermal conductivity 
of porous rock, and $Q_e$ is the source/sink term for heat extraction.

\subsection{Physics-based conceptual model:~EGS reservoir}
\label{SubSec:S2_PCM_EGS}
\begin{figure}
  \centering
  \includegraphics[scale=1.1,page=3,clip=true]
    {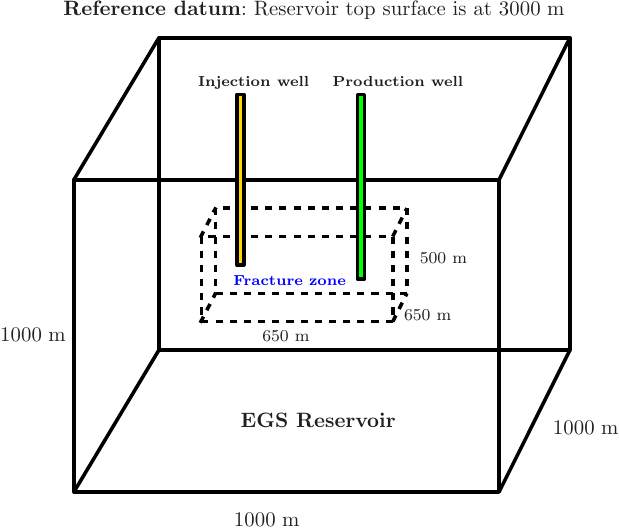}
  \caption{\textsf{Physics-based conceptual model:}~EGS reservoir 
  and fracture zone dimensions. The reservoir top surface, which 
  is the reference datum is located at 3000 m. The dimensions of 
  the reservoir are around $1000 \times 1000 \times 1000 \, 
  \mathrm{m}^3$ while the fracture zone dimensions are around 
  $650 \times 650 \times 500 \, \mathrm{m}^3$. The injection and 
  production wells are located at around $(575 \, \mathrm{m}, 575 
  \, \mathrm{m}, 450 \, \mathrm{m})$ and $(675 \, \mathrm{m}, 500 
  \, \mathrm{m}, 625 \, \mathrm{m})$.
  \label{Fig:EGS_Diagram_Phase2}}
\end{figure}

We shall briefly describe the physics-based conceptual model used 
in the numerical simulation of Phase II Fenton Hill HDR reservoir here. It should 
be noted that this conceptual model is an approximation of a more complex 
system (see the References by Kelkar et al. \cite{HDR_FentonHill_2015} and 
Brown et al. \cite{Brown_etal} for a detailed description of the reservoir). 
Such an approximation is performed to understand the essential features and 
construct a model that is amenable for numerical simulations. 
Figure \ref{Fig:EGS_Diagram_Phase2} provides a pictorial description of 
the reservoir. The reference datum, which is the reservoir top surface, 
is approximately located at a depth of $3000 \, \mathrm{m}$. The dimensions 
and volume of the reservoir are taken to be equal to $1000 \times 1000 \times 
1000 \, \mathrm{m}^3$. The fracture zone is an approximate representation of 
the region containing joint networks (fracture networks) and low-permeable 
porous rock, whose dimensions are taken to be around $650 \times 650 \times 
500 \, \mathrm{m}^3$. The fracture zone starting and ending coordinates are 
approximately taken to be $(200 \, \mathrm{m}, 200 \, \mathrm{m}, 200 \, 
\mathrm{m})$ and $(850 \, \mathrm{m}, 850 \, \mathrm{m}, 700 \, \mathrm{m})$. 
The injection and production wells are located at around $(575 \, \mathrm{m}, 
575 \, \mathrm{m}, 450 \, \mathrm{m})$ and $(675 \, \mathrm{m}, 500 \, \mathrm{m}, 
625 \, \mathrm{m})$. The distance between the wells being 215.06 $\mathrm{m}$.
These are idealized as volumetric source/sink terms in performing the numerical 
simulations. Reservoir and fracture zone porosities are assumed to be equal to 
0.0001 and 0.1. The reservoir rock density, rock specific heat capacity, rock 
thermal conductivity, rock permeability, fluid heat capacity, fluid density, 
and fluid injection temperature are taken as \mbox{2716 $\mathrm{kg} \, \mathrm{m}^
{-3}$}, \mbox{803 $\mathrm{J} \, \mathrm{kg}^{-1} \, \mathrm{K}^{-1}$}, \mbox{2.546 $\mathrm{W} 
\, \mathrm{m}^{-1} \, \mathrm{K}^{-1}$}, \mbox{$10^{-18} \, \mathrm{m}^{2}$}, \mbox{4187 $\mathrm{J} 
\, \mathrm{kg}^{-1} \mathrm{K}^{-1}$}, \mbox{950 $\mathrm{kg} \, \mathrm{m}^{-3}$}, and 
\mbox{298 $\mathrm{K}$} (Note that the above system and material parameters are taken 
from the Reference Swenson et al. \cite[Table 1]{1995_Swenson_etal_WGC}).

\begin{figure}
  \centering
  \subfigure[Injection pressure and production backpressures]
    {\includegraphics[scale=0.5]
    {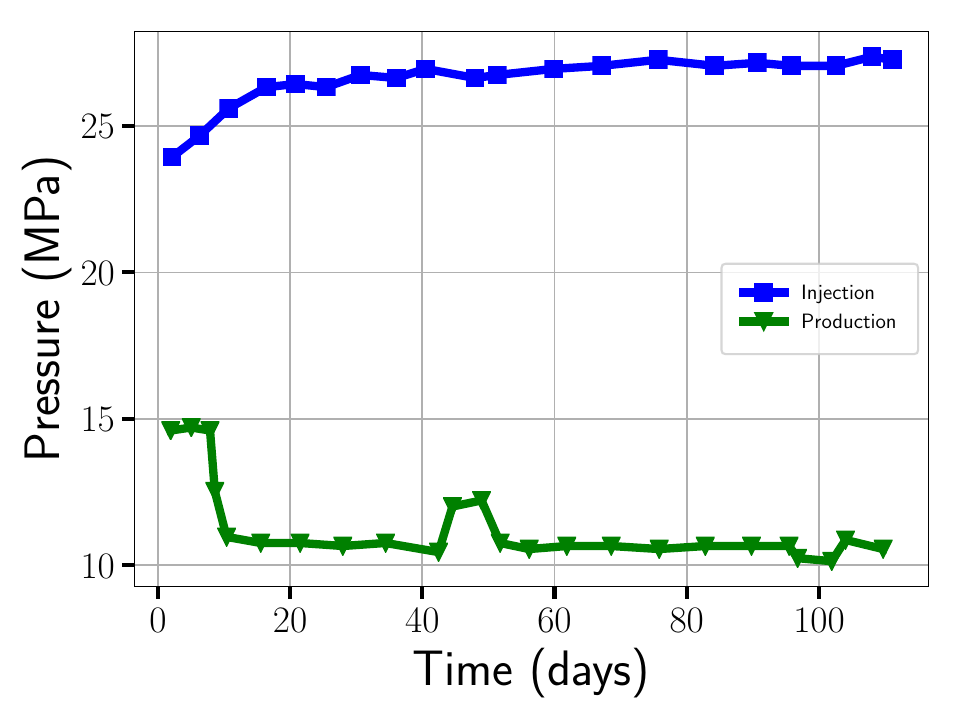}}
  \subfigure[Injection and production mass flow rates]
    {\includegraphics[scale=0.5]
    {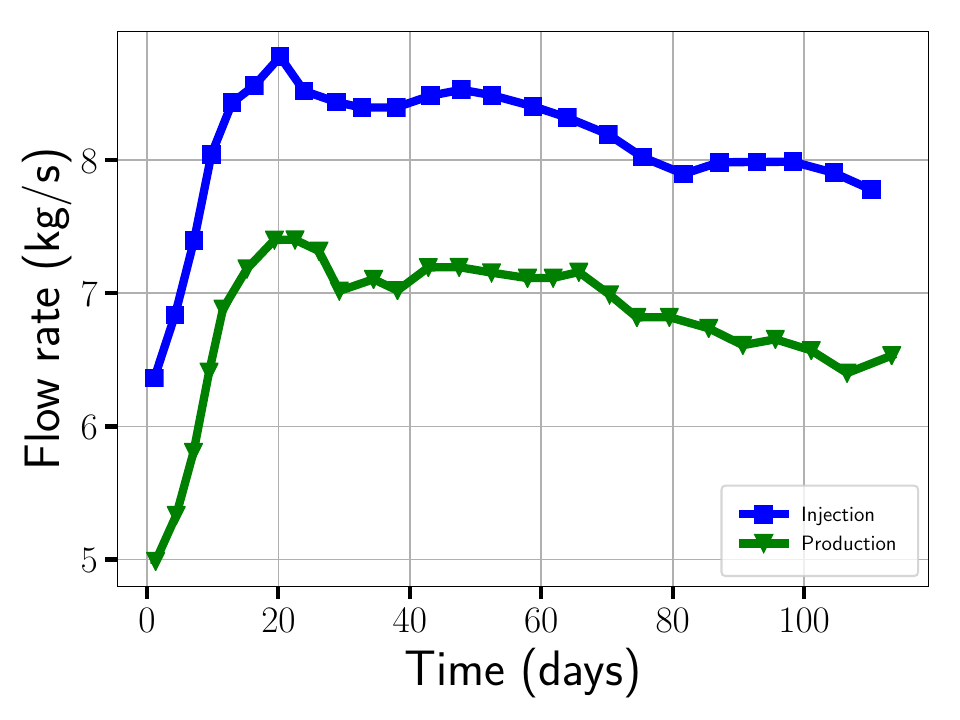}}
  \subfigure[Injection and production temperatures]
    {\includegraphics[scale=0.5]
    {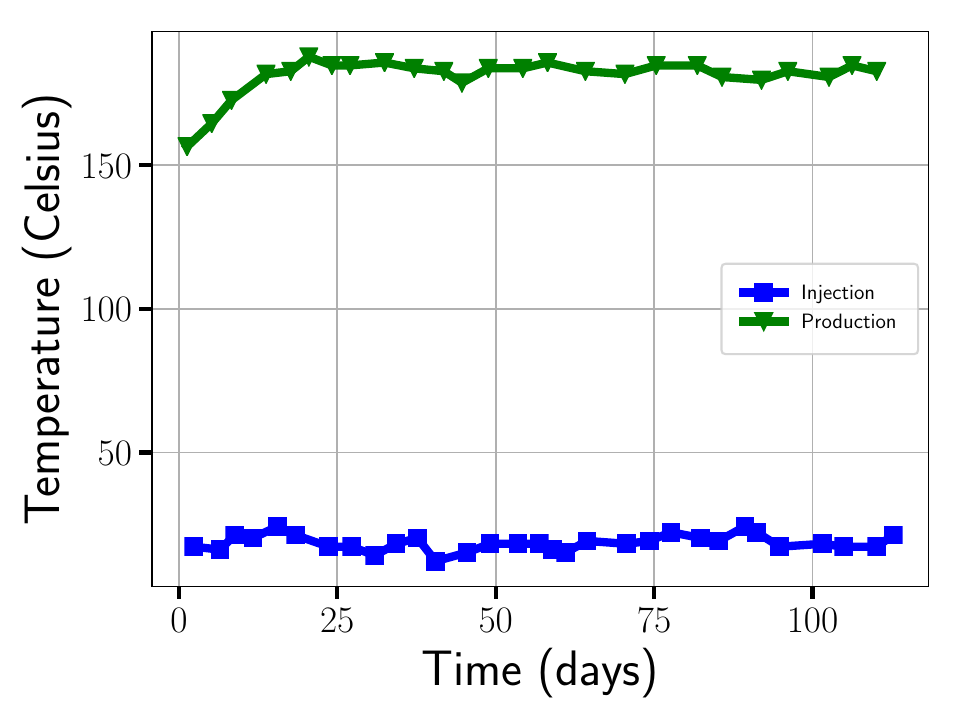}}
  \subfigure[Net thermal power output]
    {\includegraphics[scale=0.5]
    {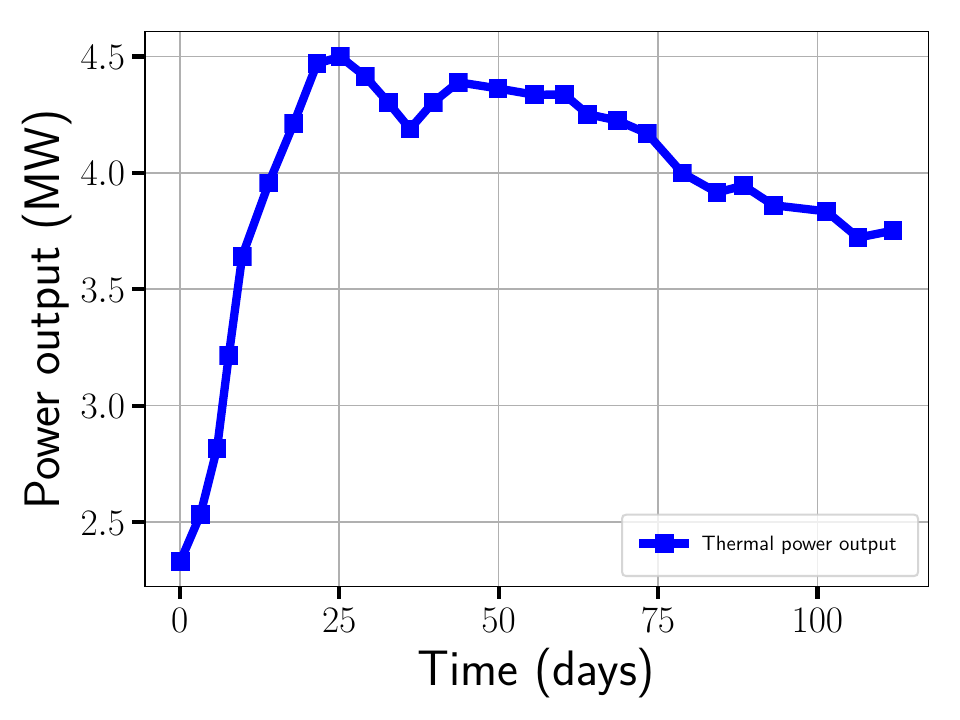}}
  \caption{\textsf{LTFT field-scale experiments of Fenton Hill 
    Phase II reservoir:}~The top left figure shows the injection 
    pressure and production backpressures (MPa) as a function of 
    time (days). The injection backpressure ranges from 25 MPa to 
    30 MPa while the production backpressure ranges from 8 MPa to 
    13 MPa. The top right figure shows the injection and production 
    mass flow rates ($\mathrm{kg} \, \mathrm{s}^{-1}$) as a function 
    of time (days). The injection mass flow rate ranges from 7.5 
    $\mathrm{kg} \, \mathrm{s}^{-1}$ to 8.5 $\mathrm{kg} \, \mathrm{s}
    ^{-1}$ while the production mass flow rate ranges from 5.5 $\mathrm{kg} 
    \, \mathrm{s}^{-1}$ to 7.0 $\mathrm{kg} \, \mathrm{s}^{-1}$. The bottom 
    left figure shows the injection and production temperatures (Celsius) 
    vs time (days). The injection temperature ranges from 293 $\mathrm{K}$ 
    ($20^{\mathrm{o}} \, \mathrm{C}$) to 303 $\mathrm{K}$ ($30^{\mathrm{o}} 
    \, \mathrm{C}$) while the production temperature ranges from 438 $\mathrm{K}$ 
    to 458 $\mathrm{K}$. Correspondingly, accounting for wellbore friction losses, 
    the bottomhole temperatures at the production well are between 453 $\mathrm{K}$ 
    ($180^{\mathrm{o}} \, \mathrm{C}$) and 498 $\mathrm{K}$ ($225^{\mathrm{o}} 
    \, \mathrm{C}$) (see Kelkar et al. \cite[Subsection 3.15.1]{HDR_FentonHill_2015}). 
    The bottom right figure shows the thermal power output (MW) during the LTFT 
    experiment.
  \label{Fig:LTFT_PhaseII_FieldData}}
\end{figure}

The initial reservoir conditions for the model are at a pressure of 13.2 MPa 
and temperature of 503 K \cite[Section-3]{HDR_FentonHill_2015}. No flow
boundary conditions are assumed for solving the flow equations. Zero gradient 
boundary conditions are assumed for solving heat transfer equations. Figure 
\ref{Fig:LTFT_PhaseII_FieldData} shows the respective injection pressure 
and production backpressure, injection and production mass flow rates, injection 
and production temperatures, and thermal power extracted during the Phase II LTFT 
experiment. These field data sets are extracted from the Hot Dry Rock Final 
Report by Kelkar et al. \cite{HDR_FentonHill_2015}. To get an estimate of 
the model parameters for the LTFT set-up, the model is calibrated against 
field-data. Levenberg-Marquardt (LM) Algorithm implemented in \textsf{MATK} 
software \cite{2016_MATK} along with PFLOTRAN is used to estimate the 
parameters -- fracture zone permeability, production temperature, 
bottom hole pressure, and skin/well factor. By varying the parameters 
around the calibrated parameter values, sensitivity analysis is performed.
Details are discussed in the next subsection.

\subsection{Numerical methodology}
\label{SubSec:S2_NM_Params_Estimation}
\begin{figure}
  \centering
  \subfigure[\textbf{Case \#1:}~Constant injection flow rate]
     {\includegraphics[scale=0.45]
     {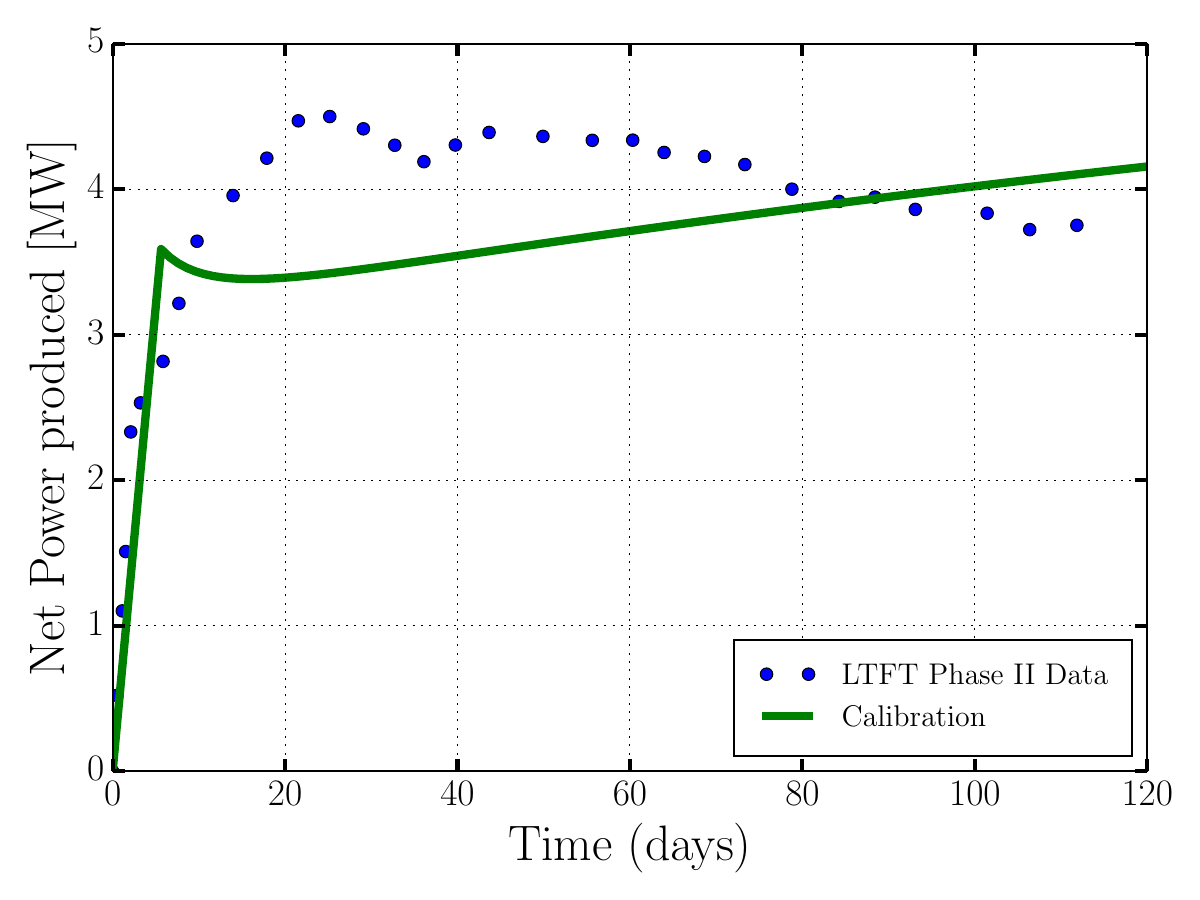}}
  \subfigure[\textbf{Case \#2:}~Time-varying injection flow rate]
     {\includegraphics[scale=0.45]
     {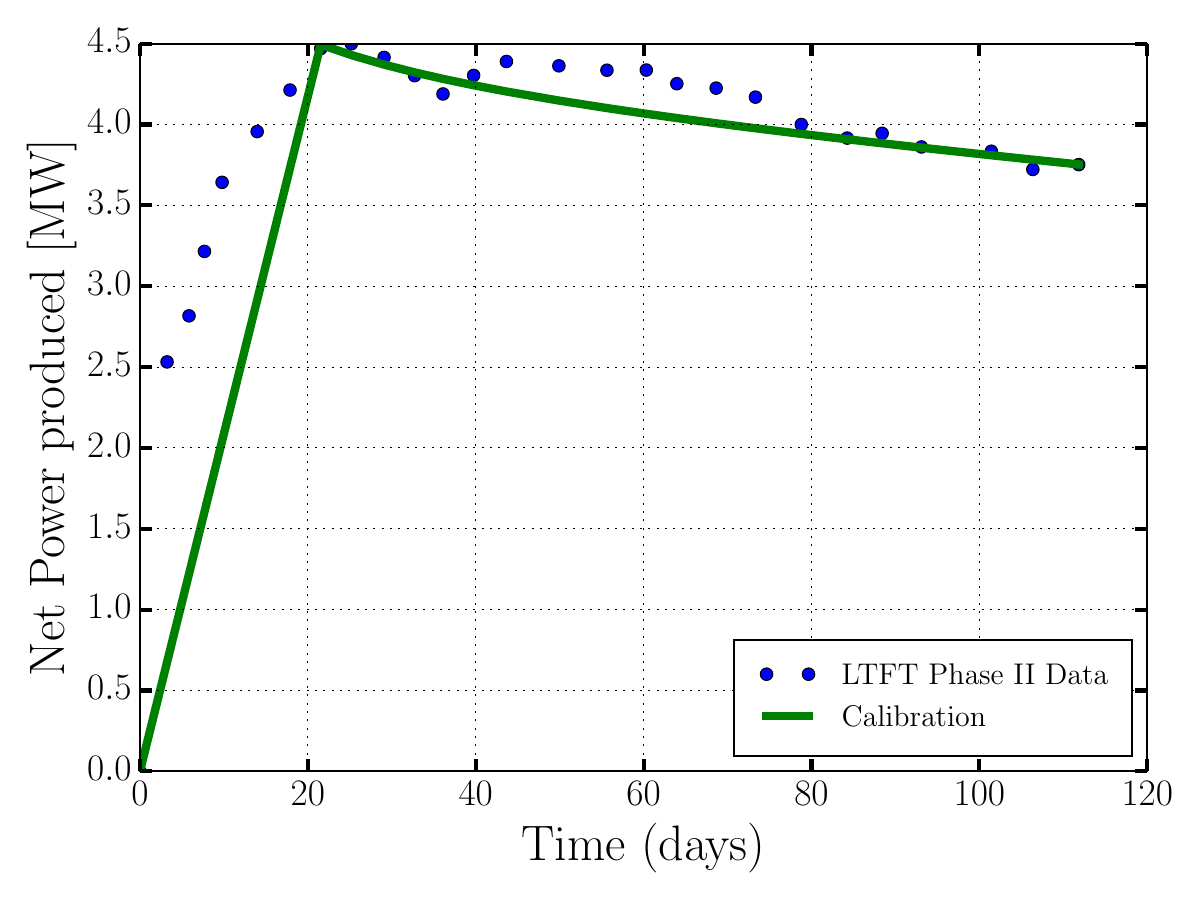}}
  \caption{\textsf{Calibration of thermal power output:}~LTFT 
    experiment and parameter estimation cases for constant 
    injection flow rate (top figure) and time-varying flow 
    rates (bottom figure). The corresponding estimated 
    parameters and calibration procedure are given in 
    Subsection \ref{SubSec:S2_NM_Params_Estimation}. 
    The time-varying injection flow 
    rates are based on Figure \ref{Fig:LTFT_PhaseII_FieldData}.
    The calibration for time-varying injection flow rate is 
    performed using Levenberg-Marquardt (LM) Algorithm implemented 
    in \textsf{MATK} software \cite{2016_MATK}. The $\mathrm{R}
    ^2$-values for constant injection and time-varying injection 
    flow rates are 0.74 and 0.68, which are almost close to each 
    other. The root mean square error (RMSE) values for each 
    calibration case is equal to 0.602 and 0.136. From these 
    RMSE values, it is evident that even though the $\mathrm{R}
    ^2$-values are close to each other the RMSE values are
    considerably different as LM algorithm calibrates by 
    minimizing the mean squared error (MSE) value. It should 
    be noted that the calibrated values for both cases are 
    of the same order.
  \label{Fig:Power_Output_Comparison}}
\end{figure}

The governing flow and heat transfer equations are solved using the PFLOTRAN 
simulator, which employs a fully implicit backward Euler for discretizing 
time and a two-point flux finite volume method for spatial discretization 
\cite[Appendix B]{2015_Lichtner_etal_PFLOTRAN}. The resulting non-linear 
algebraic equations are solved using a Newton-Krylov solver. Numerical 
simulations are performed for two different scenarios (see Figure 
\ref{Fig:Power_Output_Comparison}). Case \#1: Calibration is based on 
constant injection flow rate, which is equal to $7.5 \, \mathrm{kg} 
\, \mathrm{s}^{-1}$ \footnote{{\scriptsize It should be noted that 
the intent of LTFT experiment is to inject fluid in to reservoir 
at a constant flow rate \cite{Brown_etal,HDR_FentonHill_2015} 
so that geothermal energy could be extracted at a sustained rate. 
However, due to various operational issues, near-constant flow 
rates were observed during $20 < t < 60$ and $80 < t < 120$ days. 
There seems to be a breakthrough in injection flow rate from $0 
< t < 20$ due to wellbore breakouts \cite{Brown_etal}.}}. Case \#2: 
Calibration is performed based on time-varying injection flow rates, 
whose values are plotted in Figure \ref{Fig:LTFT_PhaseII_FieldData}. In this
case, the mean square error between thermal power output data based 
on PFLOTRAN numerical simulations and LTFT field-data is minimized using LM 
algorithm. The resulting parameters are given as follows:
\begin{itemize}
  \item [$\blacktriangleright$] Fracture zone permeability
    \begin{itemize}
      \item Constant injection:~$7.75 \times 10^{-16} \, \mathrm{m}^2$
      \item Time-varying injection:~$1.78 \times 10^{-16} \, \mathrm{m}^2$
    \end{itemize}
  \item [$\blacktriangleright$] Production temperature:~$438$ K ($165^{\mathrm{o}} 
    \mathrm{C}$), which is the same for both cases. 
  \item [$\blacktriangleright$] Bottom hole pressure:~
    \begin{itemize}
      \item Constant injection:~9.5 MPa
      \item Time-varying injection:~9.42 MPa
    \end{itemize}
  \item [$\blacktriangleright$] Skin/well factor [unit-less] to regulate mass 
    flow rate in production well:~
    \begin{itemize}
      \item Constant injection:~$3.163 \times 10^{-13}$
      \item Time-varying injection:~$5.37 \times 10^{-13}$
    \end{itemize}
\end{itemize}

The power output from the numerical simulation is calculated from the 
following expression:
\begin{align}
  \label{Eqn:Net_Power_Prod_PFLOTRAN}
  \textrm{Net power produced} &= \left( \textrm{Production mass flow rate}
  \times \textrm{Fluid heat capacity} \times 
  \textrm{Production temperature} \right) 
  \nonumber \\
  &- \left(\textrm{Injection mass flow rate} \times 
  \textrm{Fluid heat capacity} \times \textrm{Injection temperature} \right)
\end{align}

Figure \ref{Fig:Power_Output_Comparison} shows the approximate fit 
of the PFLOTRAN numerical simulation with the field-scale data of the power 
output based on the parameters estimated for Case \#1 and Case \#2. 
The $\mathrm{R}^2$-values for Case \#1 and Case \#2 are equal to 0.74 
and 0.68, which are close to each other. The mean square error (MSE) values 
for Case \#1 and Case \#2 are 0.603 and 0.018, which are considerably different. 
Reason being that the LM algorithm calibrates the model parameters by minimizing 
the MSE value. It should be noted that the calibrated parameters of the constant 
injection flow rate case and time-varying injection flow rate case are of the same 
order, which is enough for performing sensitivity analysis to identify key sensitive 
input parameters for ROMs construction. The next subsection describes the 
workflow for ROM construction and the corresponding numerical studies 
on various model parameters \cite{2015_Finsterle_WRR}.

\subsection{Sensitivity analysis, ROM development workflow, and numerical results}
\label{SubSec:S2_SA_ROM_Workflow}
In this subsection, we perform sensitivity studies on various input/output 
model parameters to model EGS reservoirs. Such an analysis is performed to 
identify key sensitive parameters for ROM inputs. These model parameters 
include mass flow rate at the injection well, skin/well factor to 
regulate mass flow rate in production well, fracture zone permeability, 
and bottom hole pressure.

\begin{figure}
  \centering
  \subfigure[Fracture zone permeability]
    {\includegraphics[scale=0.4]
    {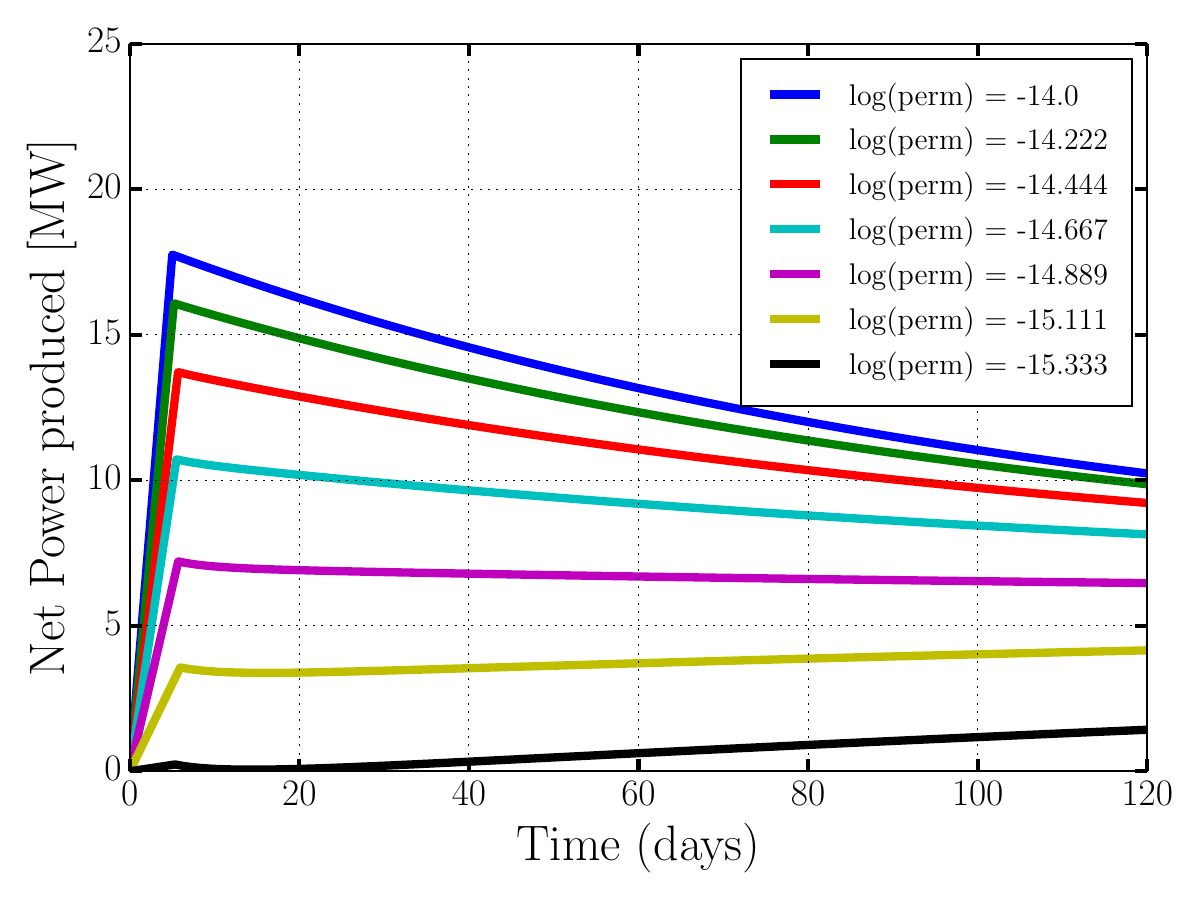}}
  \subfigure[Well/Skin factor]
    {\includegraphics[scale=0.4]
    {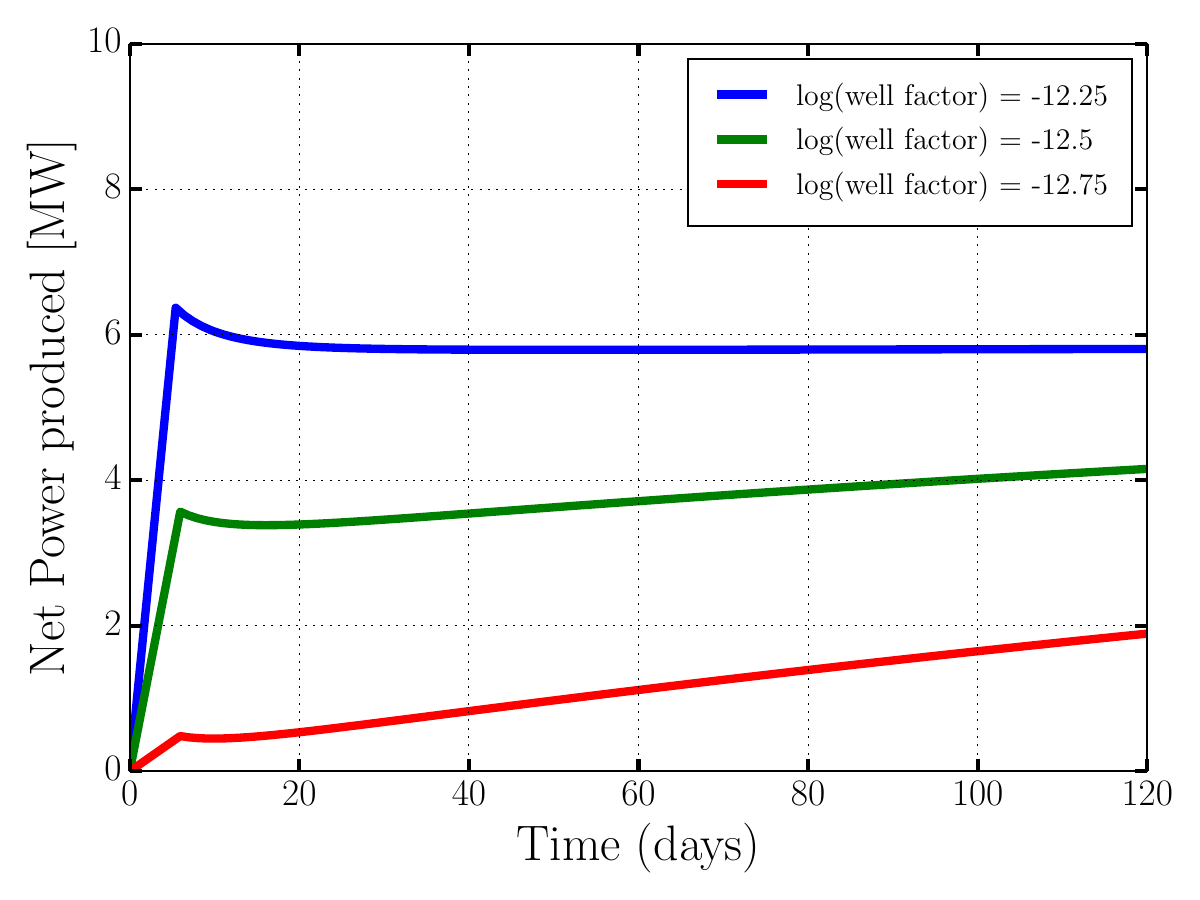}}
  \subfigure[Bottomhole pressure]
    {\includegraphics[scale=0.4]
    {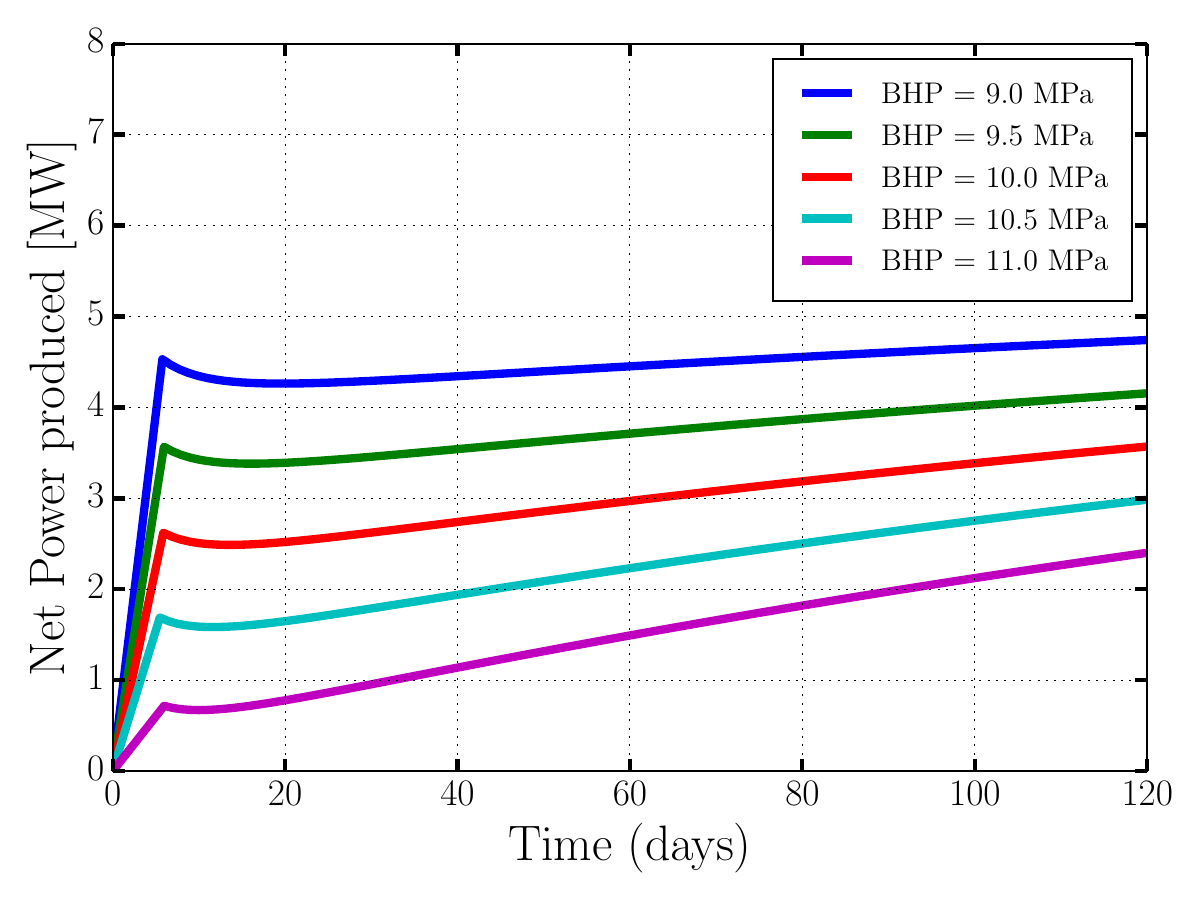}}
  \subfigure[Injection mass flow rate]
    {\includegraphics[scale=0.4]
    {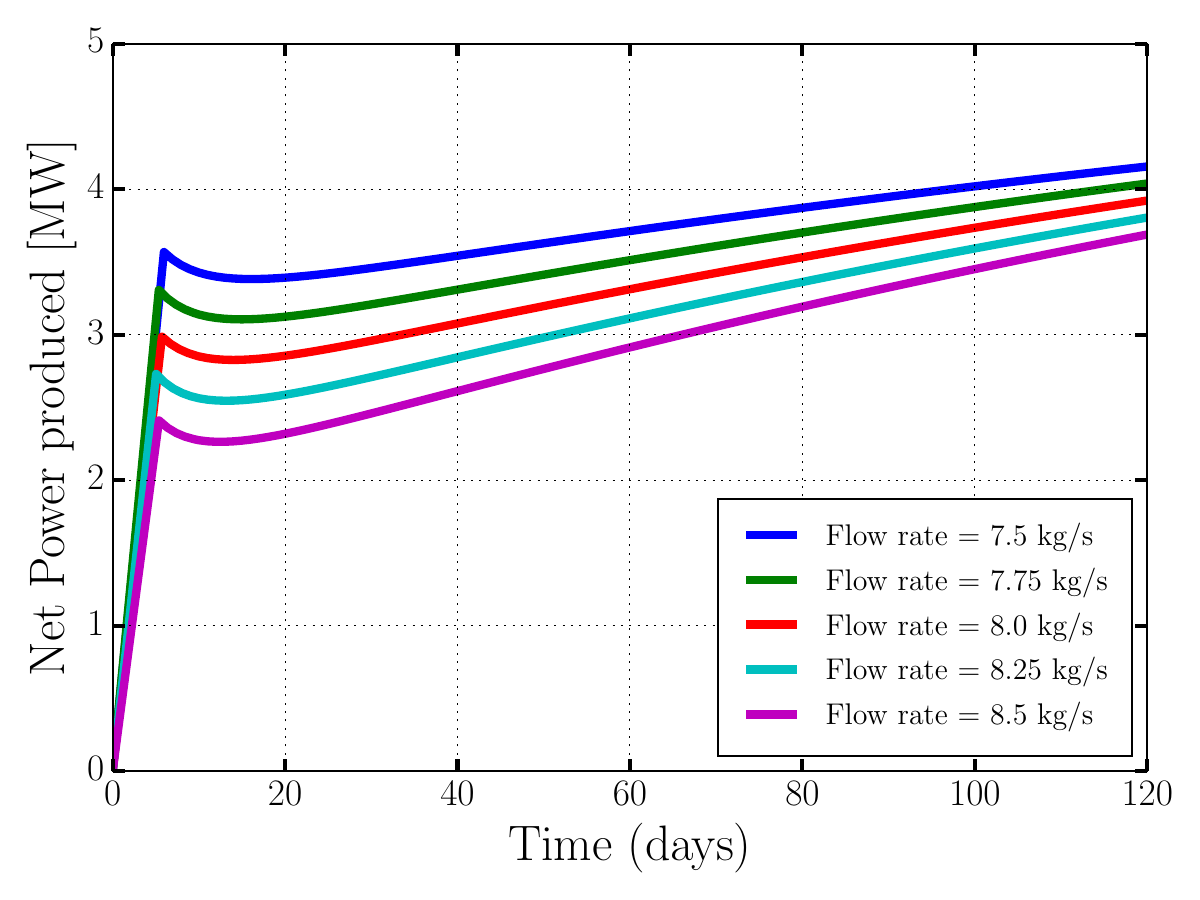}}
  \caption{\textsf{Thermal power output sensitivity studies:}~PFLOTRAN 
    simulations showing the sensitivity of thermal power production with 
    respect to fracture zone permeability (top left figure), skin/well 
    factor to regulate mass flow rate in the production well (top right 
    figure), bottom hole pressure (bottom left figure), and injection 
    mass flow rate (bottom right figure). The sensitivity analysis is 
    performed by varying fracture zone permeability, well factor, bottom 
    hole pressure, and injection mass flow rate within the range of $10^
    {-14}$ to $10^{-16}$ $\mathrm{m}^2$, $1.78 \times 10^{-13}$ to $5.62 
    \times 10^{-13}$, $9.5$ to $11$ MPa, and $7.5$ to $8.5$ $\mathrm{kg} 
    \, \mathrm{s}^{-1}$. These values are constructed based on an 
    educated guess of the fractured EGS system and considering the qualitative 
    aspects of the field-scale data (see Kelkar et al. \cite[Section-3]{HDR_FentonHill_2015}) 
    for a detailed discussion on the qualitative \& quantitative aspects of 
    Fenton Hill Phase II reservoir and corresponding approximations of LTFT 
    data). 
  \label{Fig:Sensitivity_Analysis_ROM}}
\end{figure}

\begin{figure}
  \includegraphics[scale = 0.725]{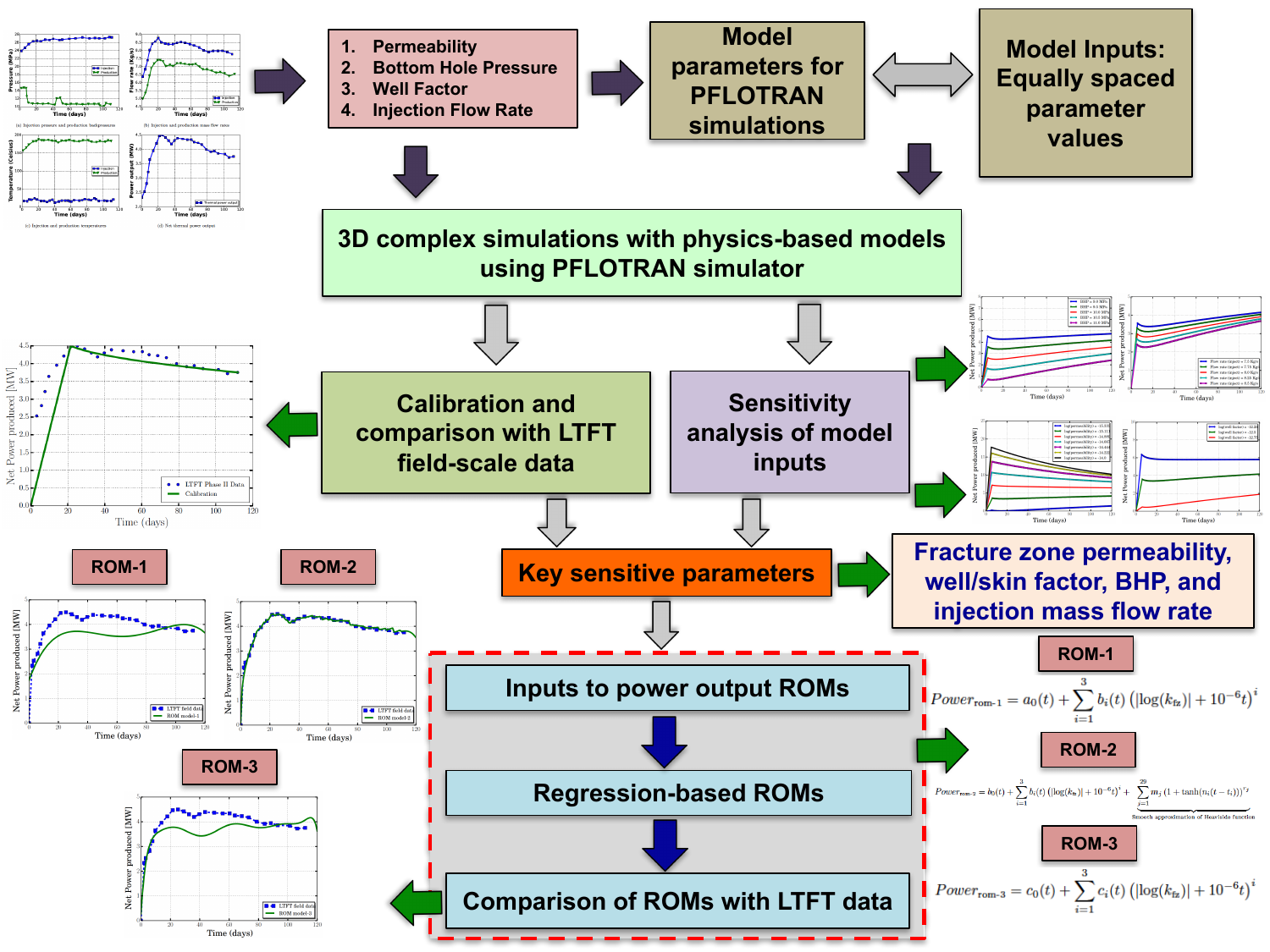}
  \caption{\textsf{ROM development flow 
    diagram for EGS reservoir:}~First, the model parameter 
    values for the PFLOTRAN simulations are constructed by 
    dividing the given parameter range into equally spaced 
    parametric values from the range. These parameters include 
    inputs from wellbore characteristics such as well factor, 
    reservoir characteristics such as fractured rock permeability, 
    bottom hole pressure, and injection flow rates. Second, 
    based on these equally spaced parameters, numerical simulations 
    are performed. Following this, sensitivity analysis is performed. 
    Key sensitive parameters are obtained from these numerical 
    sensitivity studies and the most sensitive parameter is 
    chosen as input to ROM. Herein, fractured rock permeability 
    is found to be the most sensitive parameter. The thermal power 
    output ROMs are constructed based on regression-based methods. 
    Training and validation is performed using PFLOTRAN simulation data and
    finally, the predictions from the ROMs are compared with LTFT data.
  \label{Fig:ROM_Workflow_Diagram}}
\end{figure}

Figure \ref{Fig:Sensitivity_Analysis_ROM} shows the PFLOTRAN numerical 
simulation results for the above four key parameters. These figures are 
plotted for each sensitive parameter by keeping all other parameters 
fixed to base case calibrated values. For example, if fracture zone permeability 
is varied from the base value (which is equal to $7.75 \times 10^{-16} \, 
\mathrm{m}^2$), the other three estimated values for the parameters are 
kept constant. This is done to show, identify, and rank the key sensitive 
input parameters for the ROMs. Based on Figure \ref{Fig:Sensitivity_Analysis_ROM}, 
it is apparent that the power production is highly sensitive to varying 
fracture zone permeability and least sensitive to injection mass flow rates. 
Correspondingly, the skin/well factor to control mass flow rate in production 
well and bottom hole pressure (BHP) are second and third in sensitivity ranks 
after fracture zone permeability. To consolidate, the ranking of importance 
of inputs for ROM development are given as follows (in the decreasing order):
\begin{enumerate}
  \item Fracture zone permeability (power production varies in a non-linear fashion)
  \item Skin/well factor to regulate mass flow rate in production well
  \item Bottom hole pressure
  \item Injection mass flow rate (power production varies in a linear fashion)
\end{enumerate}
ROM development is summarized as a flowchart in Figure \ref{Fig:ROM_Workflow_Diagram}.
The next section describes an approach to construct ROMs for power output.
%

\section{REDUCED-ORDER MODELING}
\label{Sec:S3_Geothermal_GE_NumModel}
The objective of model reduction methodologies is to use the knowledge generated 
by high fidelity and time-consuming numerical simulations to generate 
special functions that make use of properties of underlying systems, 
thereby obtaining a good understanding of the phenomena of interest. 
Subsection \ref{SubSec:S1_GTOCCS_ROMs} discusses many reasons why 
such a detail is warranted. We shall now provide an overview of 
some popular model order reduction methods and algorithms. These 
techniques either use physical (or other) insight or sensitivity 
studies on model parameters as a basis to reduce the complexity 
of the underlying problem and obtain a good approximation of the 
required output in an efficient way.

Some popular model reduction methods include regression-based 
model order reduction \cite{Carroll_NRAP_2014,2016_Harp_etal_IJGGC_v45_p150_p162,
2016_Keating_etal_IJGGC_v46_p187_p196}, operational model order 
reduction \cite{2008_Schilders_etal,2014_Rainieri_Fabbrocino}, 
compact reduced-order modeling \cite{2008_Schilders_etal}, truncated 
balanced realization \cite{2004_Gugercin_Antoulas_IJC_v77_p748_p766}, 
optimal Hankel-norm model order reduction \cite{2001_Antoulas_etal_CM_v280_p193_p220}, 
Gaussian process regression (GPR) \cite{2006_Rasmussen_Williams,
2013_Pau_etal_CS_v17_p705_p721}, proper orthogonal decomposition (POD) 
\cite{Buljak}, asymptotic waveform evaluation (AWE) and its variants 
\cite{Chiprout_Nakhla}, Pade via Lanczos (PVL) and its variants 
\cite{2001_Bai_Freund_LAA_v332_p139_p164}, spectral Lanczos decomposition 
method (SLDM) \cite{2002_Slone_etal_E_v22_p275_p289}, and truncation-based 
model order reduction \cite{2008_Schilders_etal}. For more details on these 
methods, algorithms, and implementation aspects (see References Qu, \cite{2004_Qu}, 
Schilders et al. \cite{2008_Schilders_etal}, Quarteroni and Rozza \cite{2014_Quarteroni_Rozza},
and Mignolet et al. \cite{2013_Mignolet_etal_JSV_v332_p2437_p2460}). Here, we 
shall construct reduced-order models based on regression-based 
model order reduction methods, which result in simple thermal power output ROMs. 
These ROMs consist of a set of algebraic relations that depend on the key sensitive 
parameters (which is fracture zone permeability) that can be evaluated very quickly.

\subsection{Reduced-order models for power output based on regression-based methods}
\label{SubSec:S3_ROMs_Interpolation}
ROMs are constructed using a combination of polynomials, 
trigonometric functions, exponential functions, and smooth approximation of 
step functions. Logistic functions are chosen as the smooth-approximation of
step functions. The rationale behind choosing such functions are as follows:
Polynomials, trigonometric functions, and exponential functions capture the 
increase and decay part of the field-scale power output data and PFLOTRAN 
numerical simulations, while the logistic functions are intended to capture 
the peaks and sudden variations. The coefficients of these functions are 
constructed through a non-linear least-squares regression fit to the PFLOTRAN 
simulations. To obtain the respective coefficients of the functions in the ROMs, 
non-linear least-squares regression was performed using the optimization solvers 
available in the open-source \textsf{Python} package \textsf{Scipy} \cite{2016_Scipy}.

Below we provide a summary of ROMs construction. 
It should be noted that all of the simulation data from the 
sensitivity analysis is used to train and validate the 
ROMs:
\begin{itemize}
  \item Training data from PFLOTRAN simulations is used to 
    construct all ROMs. These simulations are performed for 
    log(permeability) values equal to -14.0, -14.444, -14.667, 
    -14.889, and -15.333. 
  \item For ROM-2, in addition to above training datasets, a 
    small subset of field-scale data is used to construct the 
    ROM. These include the LTFT field thermal power data at 
    times $t = 0, 20, 25, 40, 60, 80, 100,$ and $120$.
  \item For validation, the PFLOTRAN simulations performed at 
    log(permeability) values equal to -14.222 and -15.111 are 
    used. 
  \item For prediction, the LTFT thermal power output field data is used. 
  \item ROM-1 is constructed using polynomials of order upto 
    degree four. ROM-2 is constructed using polynomials of 
    order upto degree eight, sine function, exponential function, 
    and smooth approximation of Heaviside functions. ROM-3 
    is constructed using polynomials of order upto degree 
    ten. 
  \item The coefficients of the ROMs are determined 
    by minimizing the sum of squares of nonlinear 
    functions. This is achieved by using \textsf{Scipy} 
    library's non-linear least-squares fit function 
    called ``\texttt{scipy.optimize.curve-fit}''.
  \item Using the developed ROMs, the $\mathrm{R}^2$-values 
    are then calculated for training, validation, and prediction 
    data.
\end{itemize}
It should be noted that a small subset of field data is 
used to train a particular ROM, which is ROM-2. At these values, we 
exactly reproduce field thermal power output for calibrated 
permeability value. However, other ROMs are not trained using 
the observation data. Observation data is first used to obtain 
realistic parameter range to generate the simulation datasets. 
These simulation datasets are then used to develop the ROMs. 
Figure \ref{Fig:Power_Output_Comparison} shows the calibration 
to obtain the base case parameter values for the simulation 
datasets. The parameters used here include fracture zone 
permeability, bottom hole pressure, and well factor. Then, 
based on a sensitivity analysis study, we found the fracture 
zone permeability to be the most sensitive parameter and so 
the ROMs were constructed as a function of time and permeability.

Next, we propose and discuss three regression-based 
reduced-order models with different levels of model parsimony.
Algorithm \ref{Algo:ROMs_Construction_Method} 
provides a detailed description of the proposed regression-based ROMs 
construction. Each model has it own pros and cons, which are 
described below:

\subsubsection{\textbf{Thermal power output ROM-1}}
\label{SubSubSec:S3_ROM1}
\begin{align}
  \label{Eqn:ROM_1}
  \textrm{Power}_{\text{\tiny {rom-1}}} = a_0(t) + \displaystyle 
  \sum \limits_{i = 1}^{3} a_i(t) \left( |\mathrm{log}(k_
  {\text{\tiny {fz}}}) | + 10^{-6} t \right)^i
\end{align}
where $t$ denotes the time in days, the coefficients $a_i$ are 
function of time (days), and $k_{\text{\tiny {fz}}}$ is the 
fracture zone permeability.

\subsubsection{\textbf{Thermal power output ROM-2}}
\label{SubSubSec:S3_ROM2}
\begin{align}
  \label{Eqn:ROM_2}
  \textrm{Power}_{\text{\tiny {rom-2}}} = b_0(t) + \displaystyle 
  \sum \limits_{i = 1}^{3} b_i(t) \left( |\mathrm{log}(k_
  {\text{\tiny {fz}}}) | + 10^{-6} t \right)^i + \underbrace{ 
  \displaystyle \sum \limits_{j = 1}^{29} m_j \left(1 + 
  \tanh(n_j (t - t_j)) \right)^{r_j}}_{\text{\tiny {Smooth 
  approximation of Heaviside function}}}
\end{align}

\subsubsection{\textbf{Thermal power output ROM-3}}
\label{SubSubSec:S3_ROM3}
\begin{align}
  \label{Eqn:ROM_3}
  \textrm{Power}_{\text{\tiny {rom-3}}} = c_0(t) + \displaystyle 
  \sum \limits_{i = 1}^{3} c_i(t) \left( |\mathrm{log}(k_
  {\text{\tiny {fz}}}) | + 10^{-6} t \right)^i
\end{align}

\begin{algorithm} 
  \caption{{\small A numerical methodology to construct regression-based 
    reduced-order models for EGS}}
  \label{Algo:ROMs_Construction_Method} 
  \begin{algorithmic}[1]
    \STATE INPUTs for PFLOTRAN simulations:~Fracture zone permeability, 
      bottom hole pressure, well factor, and injection flow rate.
    \STATE To construct regression-based ROMs, get high-fidelity 
      numerical simulation data by running PFLOTRAN simulator for 
      different parameter values. Total number of simulations = 2625.
      \begin{itemize}
         \item Logarithm of fracture zone permeability = $-14.0, -14.222, 
           -14.444, -14.667,$ \newline $-14.889, -15.111, -15.333$.
         \item Logarithm of well factor = $-12.25, -12.5, -12.75$.
         \item Bottom hole pressure = $9.0, 9.5, 10.0, 10.5, 11.0$.
         \item Injection mass flow rate = $7.5, 7.75, 8.0, 8.25, 8.5$.
      \end{itemize}
    \STATE Post-process each high-fidelity numerical simulation to obtain 
      thermal power output as function of time. Then, perform sensitivity 
      analysis and identify the dominant parameter. In our case, we found 
      fracture zone permeability is dominant. See Subsection \ref{SubSec:S2_SA_ROM_Workflow} 
      for more details.
    \STATE The proposed regression-based ROMs are function of time and 
      dominant parameter, which is fracture zone permeability. See 
      Equations \eqref{Eqn:ROM_1}--\eqref{Eqn:ROM_3}.
    \STATE To construct, train, and validate ROMs, among 2625 high-fidelity 
      simulation dataset we choose a subset with varying fracture zone permeability 
      (while other parameters are kept constant). The values of the other 
      parameters correspond to the calibration case. 
      \begin{itemize}
        \item Calibration case parameter values are: Logarithm of well 
          factor = $-12.5$, bottom hole pressure = $9.5$, and injection 
          mass flow rate = $7.5$.
        \item There are a total of 7 PFLOTRAN simulations that are used in 
          the proposed ROM construction methodology. Out of 7 simulations, 
          5 simulations are used for training and 2 simulations are used 
          for validation of ROMs. LTFT thermal power output data is used 
          for prediction.
      \end{itemize}
     \STATE ROM-1 construction:~{\scriptsize $a_0(t) + \displaystyle 
       \sum \limits_{i = 1}^{3} a_i(t) \left( |\mathrm{log}(k_{
	   \text{\tiny {fz}}}) | + 10^{-6} t \right)^i$}
       \begin{itemize}
         \item Polynomials of order upto degree four are selected.
         \item The coefficients in $a_i(t)$ are obtained my minimizing 
           the error between high-fidelity simulation data and the explicit 
           expression of ROM-1.
         \item The coefficient values, which are obtained by solving 
            the non-linear least-squares regression using Levenberg-Marquardt 
            algorithm are given by Equations \eqref{Eqn:ROM1_Coeff0}--\eqref{Eqn:ROM1_Coeff3}.
       \end{itemize}
     \STATE ROM-2 construction:~{\scriptsize $b_0(t) + \displaystyle 
       \sum \limits_{i = 1}^{3} b_i(t) \left( |\mathrm{log}(k_{\text{\tiny {fz}}})| 
       + 10^{-6} t \right)^i + \displaystyle \sum \limits_{j = 1}^{29} m_j \left(1 
       + \tanh(n_j (t - t_j)) \right)^{r_j}$}
       \begin{itemize}
         \item Polynomials of order upto degree eight, sine, exponential, 
           and smooth approximation of Heaviside functions are selected.
         \item The non-linear least-squares regression problem is solved 
           with the constraint that at times $t = 0, 20, 25, 40, 60, 80, 
           100,$ and $120$, the ROM-2 model output matches the LTFT thermal 
           power output data. The coefficients are given by Equations 
           \eqref{Eqn:ROM2_Coeff0}--\eqref{Eqn:ROM2_Param11}.
       \end{itemize}
     \STATE ROM-3 construction:~{\scriptsize $a_0(t) + \displaystyle 
       \sum \limits_{i = 1}^{3} a_i(t) \left( |\mathrm{log}(k_{
	   \text{\tiny {fz}}}) | + 10^{-6} t \right)^i$}
       \begin{itemize}
         \item Polynomials of order upto degree ten are selected. 
           ROM-3 is constructed in similar fashion to ROM-1. The 
           coefficient values in $c_i(t)$ are given by Equations 
           \eqref{Eqn:ROM3_Coeff0}--\eqref{Eqn:ROM3_Coeff3}.
       \end{itemize}
     \STATE OUTPUTS:~Regression models for ROM-1, ROM-2, and ROM-3. Model 
       expressions are given in Appendix. They are used to predict thermal 
       power output.
  \end{algorithmic}
\end{algorithm}

\begin{table}
  \centering
	\caption{Summary of proposed reduced-order models 
	  for thermal power output. A small subset of field-scale data 
	  is used to construct the ROM-2. These include the LTFT field 
	  thermal power data at times $t = 0, 20, 25, 40, 60, 80, 100,$ 
	  and $120$.
	\label{Tab:ROMs_Summary_1}}
	\begin{tabular}{|c|c|} \hline
	  \multirow{2}{*}{{\scriptsize Model}} & 
	  \multirow{2}{*}{{\scriptsize Thermal Power Output}} \\
	  & \\ \hline
	  {\scriptsize ROM-1} & {\scriptsize $a_0(t) + \displaystyle \sum 
	  \limits_{i = 1}^{3} a_i(t) \left( |\mathrm{log}(k_{
	  \text{\tiny {fz}}}) | + 10^{-6} t \right)^i$}  \\
	  {\scriptsize ROM-2} & {\scriptsize $b_0(t) + \displaystyle 
      \sum \limits_{i = 1}^{3} b_i(t) \left( |\mathrm{log}(k_{\text{\tiny 
      {fz}}}) | + 10^{-6} t \right)^i + \displaystyle \sum \limits
      _{j = 1}^{29} m_j \left(1 + \tanh(n_j (t - t_j)) \right)^{r_j}$} \\
	  {\scriptsize ROM-3} & {\scriptsize $c_0(t) + \displaystyle \sum \limits_
	  {i = 1}^{3} c_i(t) \left( |\mathrm{log}(k_{\text{\tiny {fz}}}) | 
	  + 10^{-6} t \right)^i$} \\
	  \hline
	\end{tabular}
\end{table}

\begin{table}
  \centering
	\caption{$R^2$-values for training, validation, and 
	  prediction of numerical simulations and LTFT field-data. Out 
	  of total seven simulation datasets, 5 were used for training 
	  and 2 for validation.
	  To construct training dataset, numerical simulations 
	  are performed for log(permeability) values equal to 
	  -14.0, -14.444, -14.667, -14.889, and -15.333. To 
	  construct validation dataset, numerical simulations 
	  are performed for log(permeability) values of -14.222 
	  and -15.111. Prediction dataset being the LTFT thermal 
	  power output field-data.
	\label{Tab:ROMs_Summary_2}}
	\begin{tabular}{|c|c|c|c|} \hline
	  \multirow{2}{*}{{\scriptsize Model}} & 
      \multicolumn{3}{|c|}{{\scriptsize $R^2$-values}} \\ 
      \cline{2-4}
	  & {\scriptsize Training} & {\scriptsize Validation} & {\scriptsize 
	  Prediction} \\ \hline
	  {\scriptsize ROM-1} & {\scriptsize 
	  0.489, 0.326, 0.22, 0.75, and 0.96} & {\scriptsize 0.406 and 0.407}
	  & {\scriptsize 0.668} \\
	  {\scriptsize ROM-2} & {\scriptsize 0.877, 0.804, 0.796, 0.689, and 
	  0.573} & {\scriptsize 0.817 and 0.542} & {\scriptsize 0.986}  \\
	  {\scriptsize ROM-3} & {\scriptsize 0.917, 0.896, 0.9, 0.889, 
	  and 0.813} & {\scriptsize 0.892 and 0.893} & {\scriptsize 0.824}  \\
	  \hline
	\end{tabular}
\end{table}

The above ROMs are used to predict LTFT thermal power output 
data as function of time and fracture zone permeability. As per computational 
cost, the time taken to run a PFLOTRAN simulation is around 45 seconds. 
To construct the regression ROMs described above we need 5 high-fidelity 
simulations for training and 2 high-fidelity simulations for validation.
Hence, the total time taken to run 7 high-fidelity simulations is around 
315 seconds. To construct ROM-1, ROM-2, and ROM-3, we need to find coefficients 
of $a_i(t)$, $b_i(t)$, and $c_i(t)$. The coefficients of the ROMs are obtained 
by minimizing the sum of squares of nonlinear functions with respect to training 
data using Levenberg-Marquardt algorithm. The procedure to obtain these coefficients 
is described in Algorithm \ref{Algo:ROMs_Construction_Method} and the coefficient 
values are given in Appendix. The Levenberg-Marquardt algorithm takes around 0.06, 
0.16, and 0.1 seconds to obtain the coefficients of ROM-1, ROM-2, and ROM-3. The 
time taken by ROM-1, ROM-2, and ROM-3 to make a prediction is around 0.0002, 0.0015, 
and 0.0008 seconds. This means the ROMs are $10^4$ times faster than a high-fidelity 
numerical simulation.

\begin{remark}
  The regression-based ROMs are specific to 
  this EGS application. The proposed ROMs can be generalized 
  by removing site-specific aspects in ROMs construction in 
  Algorithm \ref{Algo:ROMs_Construction_Method}. These include, 
  relaxing the range of injection rates, well factor values, 
  bottom hole pressures, and fracture zone permeability. For 
  a new site, first we need to constrain the parameter space 
  specific to that site. Then, we just have to use the same 
  methodology proposed in Algorithm \ref{Algo:ROMs_Construction_Method} 
  to construct regression-based ROMs specific to the new site.
\end{remark}

\subsection{Discussion and inferences: Predictive capabilities of ROMs with respect 
  to field-scale data and PFLOTRAN}
\label{SubSec:S3_ROMs_Discussion}
\begin{figure}
  \centering
  \subfigure[\textbf{Training:} PFLOTRAN simulations 
     and ROM-1 ]
    {\includegraphics[scale=0.4]
    {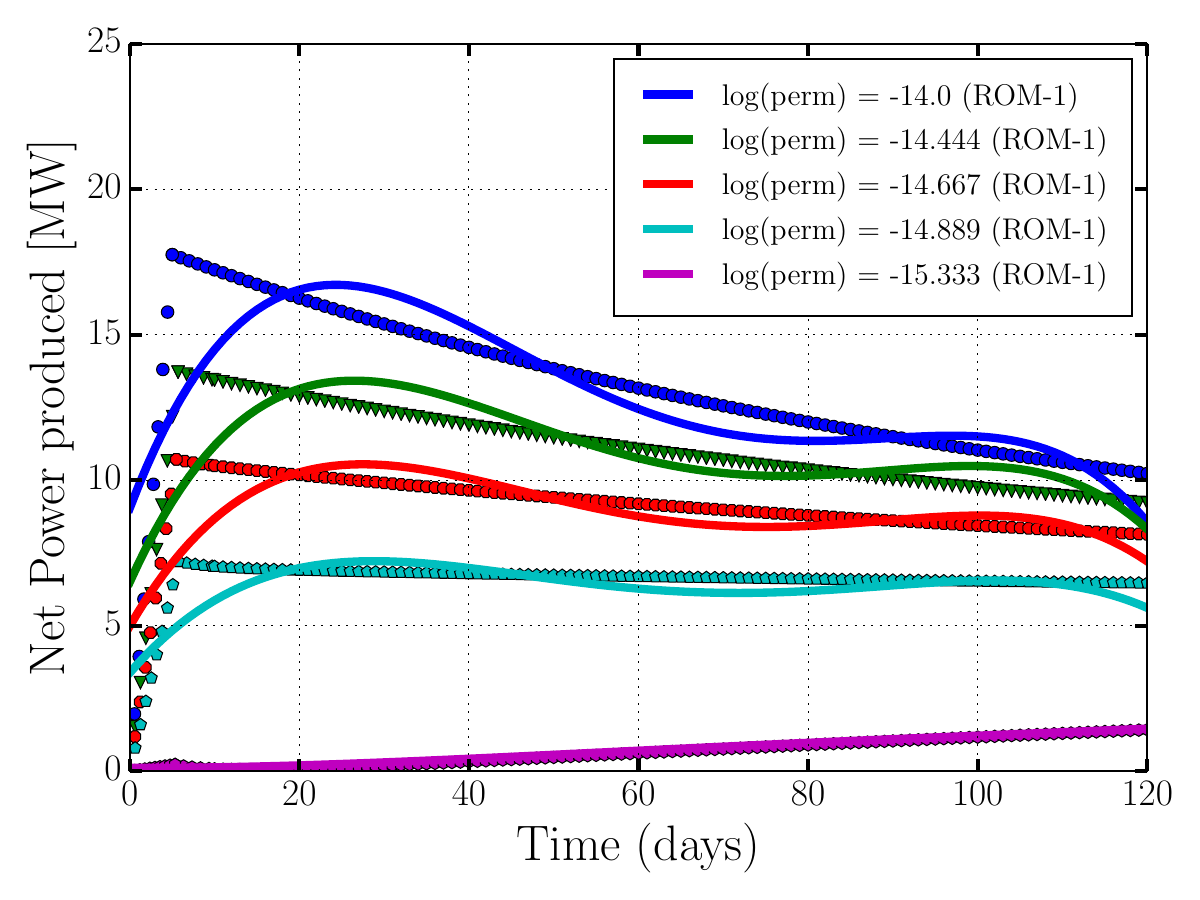}}
  \subfigure[\textbf{Validation:} PFLOTRAN simulations 
    and ROM-1 ]
    {\includegraphics[scale=0.4]
    {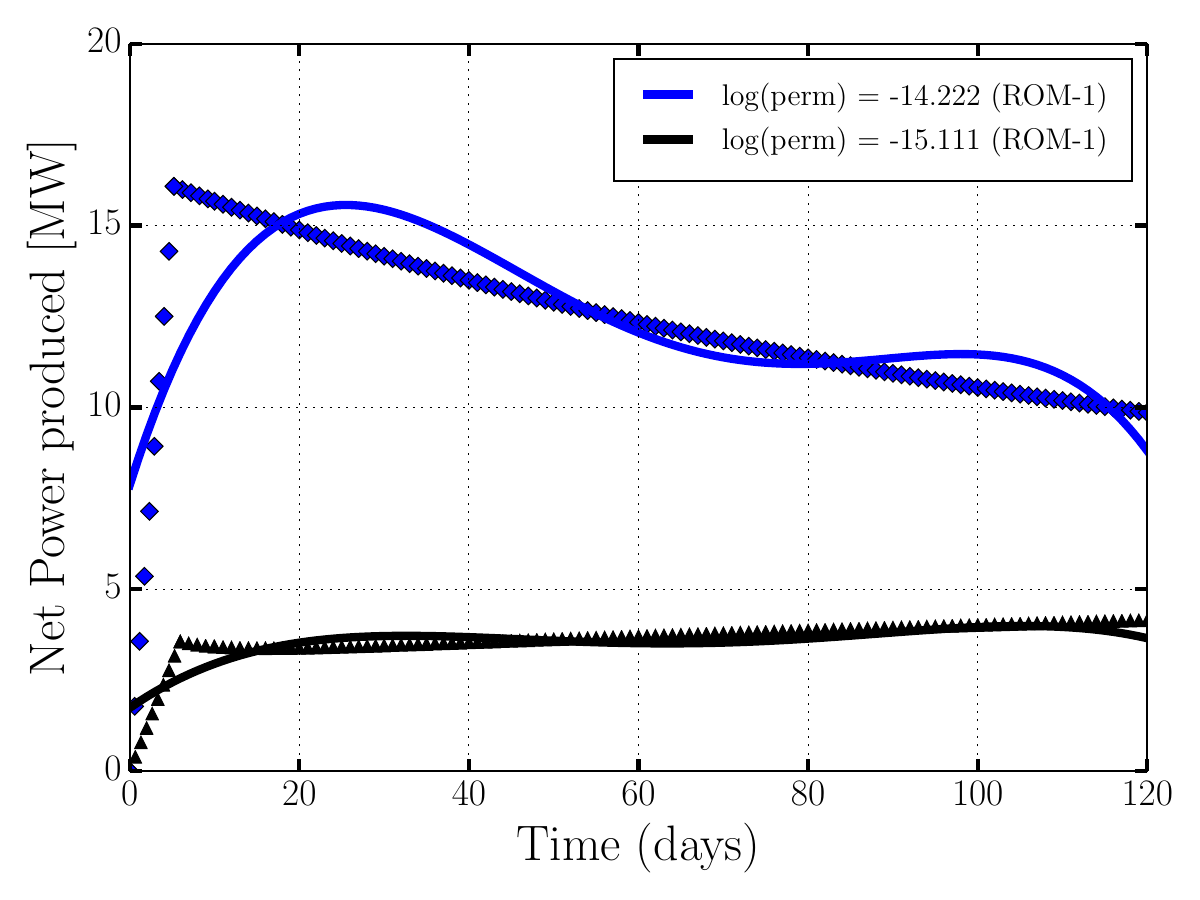}}
  \subfigure[\textbf{Prediction:} Comparison of LTFT power data and ROM-1]
    {\includegraphics[scale=0.4]
    {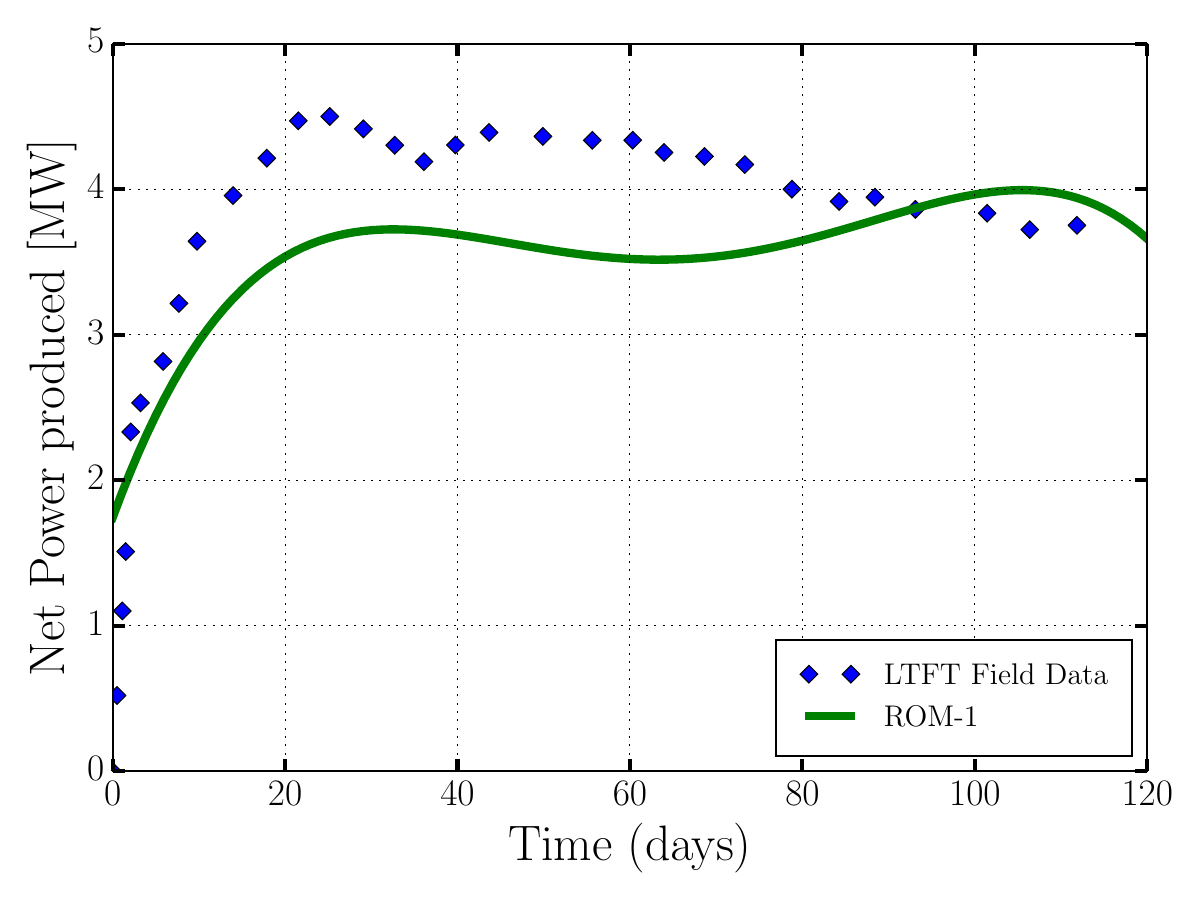}}
  \caption{\textsf{Thermal power output ROM-1:}~The top 
    left figure compares the ROM-1 output with PFLOTRAN numerical 
    simulations over a training set. Numerical simulation data is shown with markers and 
    ROMs are shown with solids of same color. 
    For training, the $\mathrm{R}^2$-values 
    are equal to 0.489, 0.326, 0.22, 0.75, and 0.96 for log(perm) values 
    of -14.0, -14.444, -14.667, -14.889, and -15.333. The top right figure 
    is the validation of the proposed ROM-1 with the PFLOTRAN simulations. 
    For validation, the $\mathrm{R}^2$-values are equal to 0.406 and 0.407 
    for log(perm) values of -14.222 and -15.111. The entire data set consists 
    of 7 numerical simulations out of which 5 simulations were used for training 
    and 2 simulations were used for validation of the developed ROM-1. The bottom 
    figure compares the predictions of ROM-1 with LTFT field-scale power output 
    data set of Phase II experiment. The fracture zone permeability used is
    the calibrated value with constant injection flow rate. The 
    $\mathrm{R}^2$-value for this prediction is equal to 0.668. From this figure, 
    it can be concluded that ROM-1 is able to accurately reproduce the power output 
    of numerical simulations for certain low values of permeability. However, as the 
    fracture zone permeability increases, there is a considerable deviation between 
    ROM-1 outputs and PFLOTRAN simulations.
  \label{Fig:PFLOTRAN_LTFTdata_ROM1}}
\end{figure}

We shall now provide a rationale behind the construction of these ROMs. 
Moreover, we shall analyze the capabilities of these three ROMs
in describing the trends in field-scale data and PFLOTRAN simulations. 
The construction of ROM-1 is purely based on polynomials. This ROM
consists of power-series terms (up to order four) involving natural logarithm 
of fracture zone permeability and time. The coefficients of the ROM-1 are constructed 
by matching the power output of PFLOTRAN numerical simulations at certain fixed intervals 
of time (for various fracture zone permeabilities). Figure \ref{Fig:PFLOTRAN_LTFTdata_ROM1} 
shows the training, validation using numerical simulations of ROM-1 along with its predictions. The
predictions are compared with field power data.
The behavior of ROM-1 is clearly distinct from the PFLOTRAN predictions (i.e., PFLOTRAN 
predicts a rapid increase in the first few days followed by a smooth decline over the 
rest of the time period, whereas ROM-1 shows a more gradual rise followed by a varying 
decline).  Nevertheless, for time periods between 20-100 days, ROM-1 is able to accurately 
reproduce the power output of numerical simulations for low values of fracture zone 
permeability. However, as the fracture zone permeability increases, there is a considerable 
deviation between ROM-1 outputs and PFLOTRAN numerical simulations. This is because ROM-1 is 
constructed by matching PFLOTRAN numerical simulations only at certain time intervals. 
Moreover, the polynomial order considered to construct the ROM is very low. 

\begin{figure}
  \centering
  \subfigure[\textbf{Training:} PFLOTRAN 
    simulations and ROM-2 ]
    {\includegraphics[scale=0.4]
    {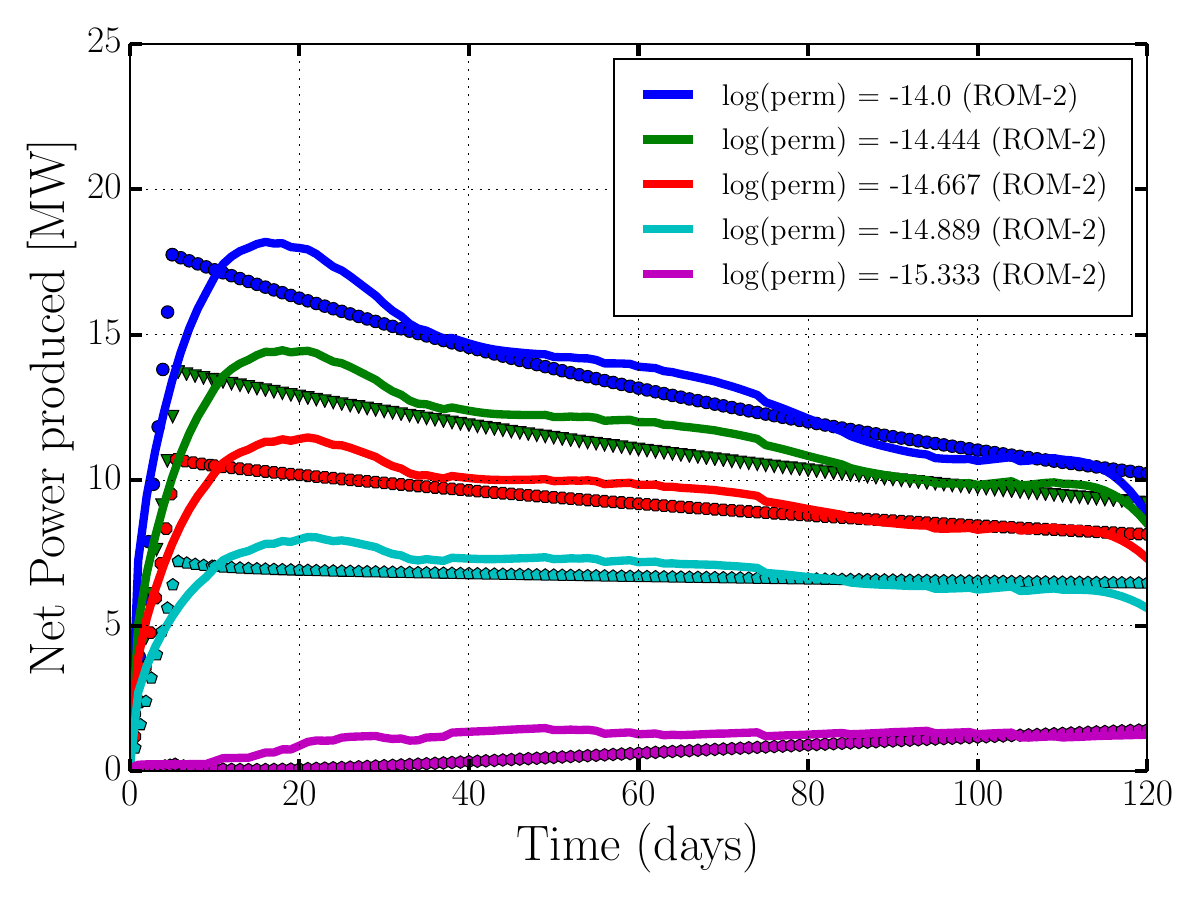}}
  \subfigure[\textbf{Validation:} PFLOTRAN 
    simulations and ROM-2 ]
    {\includegraphics[scale=0.4]
    {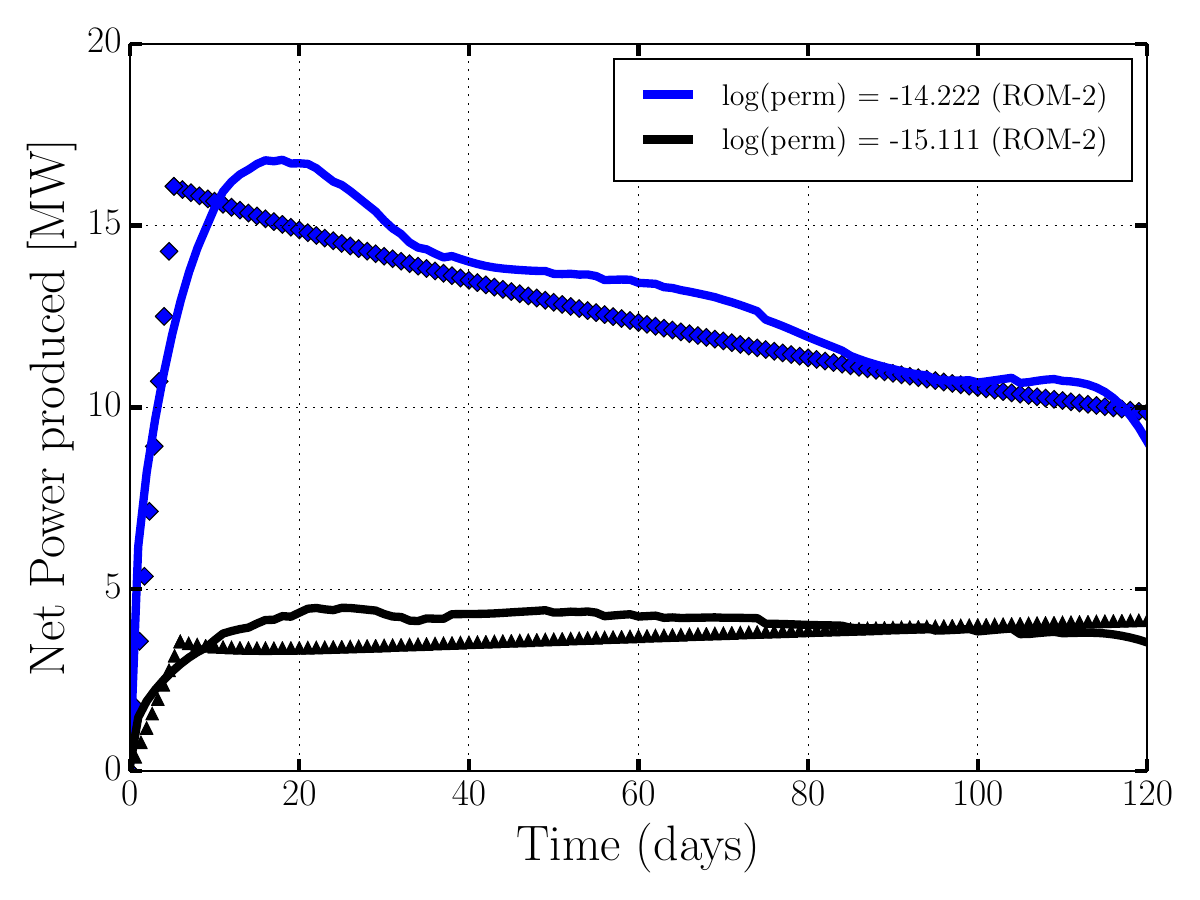}}
  \subfigure[\textbf{Prediction:} Comparison of LTFT 
    power data and ROM-2]
    {\includegraphics[scale=0.4]
    {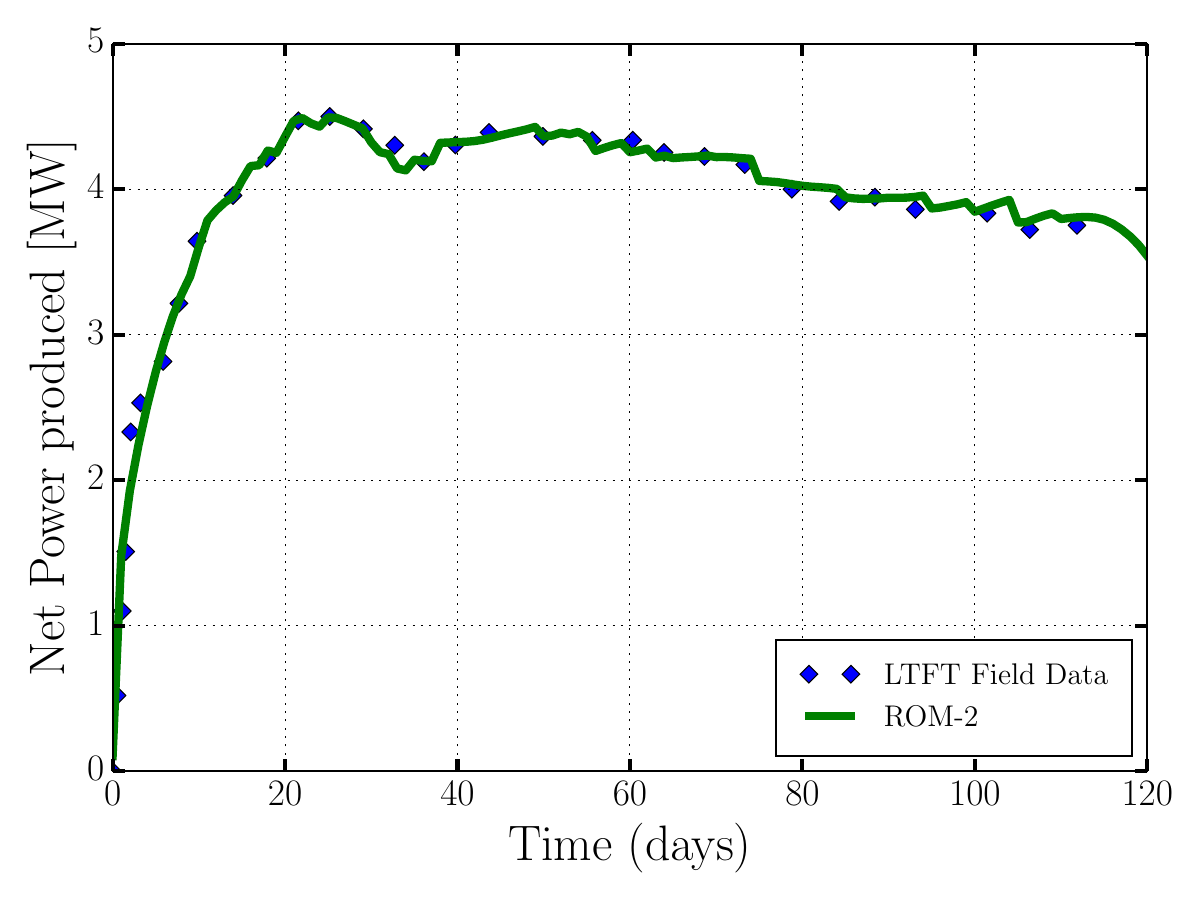}}
  \caption{\textsf{Thermal power output ROM-2:}~The top 
    left figure compares the ROM-2 output with PFLOTRAN numerical 
    simulations over a training set.  Numerical simulation data is shown with markers and 
    ROMs are shown with solids of same color. 
    For training, the $\mathrm{R}^2$-values 
    are equal to 0.877, 0.804, 0.796, 0.689, and 0.573 for log(perm) values 
    of -14.0, -14.444, -14.667, -14.889, and -15.333. The top right figure 
    is the validation of the proposed ROM-2 with the PFLOTRAN simulations. 
    For validation, the $\mathrm{R}^2$-values are equal to 0.817 and 0.542 
    for log(perm) values of -14.222 and -15.111. The entire data set consists 
    of 7 numerical simulations out of which 5 simulations are used for training 
    and 2 simulations are used for validation of the developed ROM-2. The bottom 
    figure compares the predictions of ROM-2 with LTFT field-scale power output 
    data set of Phase II experiment. The fracture zone permeability 
    used is calibrated value for constant injection flow rate. The 
    $\mathrm{R}^2$-value for this prediction is equal to 0.986. From this figure, 
    the following can be concluded:~ROM-2 reproduces the PFLOTRAN simulations 
    better than ROM-1 at higher permeabilities. However, for low permeabilities, 
    there is a considerable deviation between ROM-2 outputs and PFLOTRAN 
    simulations. Interestingly, ROM-2 outputs closely matches the LTFT experiment, 
    qualitatively and quantitatively.
  \label{Fig:PFLOTRAN_LTFTdata_ROM2}}
\end{figure}

In terms of predicting the field-scale data, ROM-1 is able to reproduce only certain 
qualitative features of the field-scale data (Figure \ref{Fig:PFLOTRAN_LTFTdata_ROM1}) 
and is relatively parsimonious. These aspects include the initial increase of power 
output, the corresponding decrease after the time $t = 20$ days, and then an increase 
in the power output after $t = 100$ days. However, quantitatively, the difference in 
power output values of ROM-1 and LTFT experiment are high. Figure 
\ref{Fig:PFLOTRAN_LTFTdata_ROM2} shows the behavior of ROM-2. 
From equation \eqref{Eqn:ROM_2}, it is evident that ROM-2 has more 
number of terms and coefficients than ROM-1. The motivation behind the construction 
of such a model is that we would like to accurately describe the LTFT experiment at 
various time intervals. ROM-2 is constructed by adding smooth approximations of step 
functions and higher-order polynomials to ROM-1. Figure \ref{Fig:PFLOTRAN_LTFTdata_ROM2} 
shows the training and validation against numerical simulations and final predictions of ROM-2.
The predictions are compared with LTFT experiment. From this figure, it evident that ROM-2 shares some of the limitations 
of ROM-1:~It does not reproduce the rapid rise in net power produced for the initial 
days of operation (albeit the fit is better than ROM-1), nor does it reproduce a nearly 
linear decline for times out to 120 days. ROM-2 qualitatively reproduces the decline 
portion of the PFLOTRAN predictions (for days 20-120), but it over predicts the numerical 
simulations, quantitatively.  In general, ROM-2 reproduces the PFLOTRAN results better 
than ROM-1 at higher permeabilities. Interestingly, power output predicted using ROM-2 
closely matches the LTFT experiment, qualitatively and quantitatively. However, 
ROM-2 is overfitted (see Table \ref{Tab:ROMs_Summary_2} for $\mathrm{R}^2$-values). In terms of reproducing 
the field-scale data and PFLOTRAN simulations at higher permeability ROM-2 is certainly 
better than ROM-1.

\begin{figure}
  \centering
  \subfigure[\textbf{Training:} PFLOTRAN 
    simulations and ROM-3 ]
    {\includegraphics[scale=0.4]
    {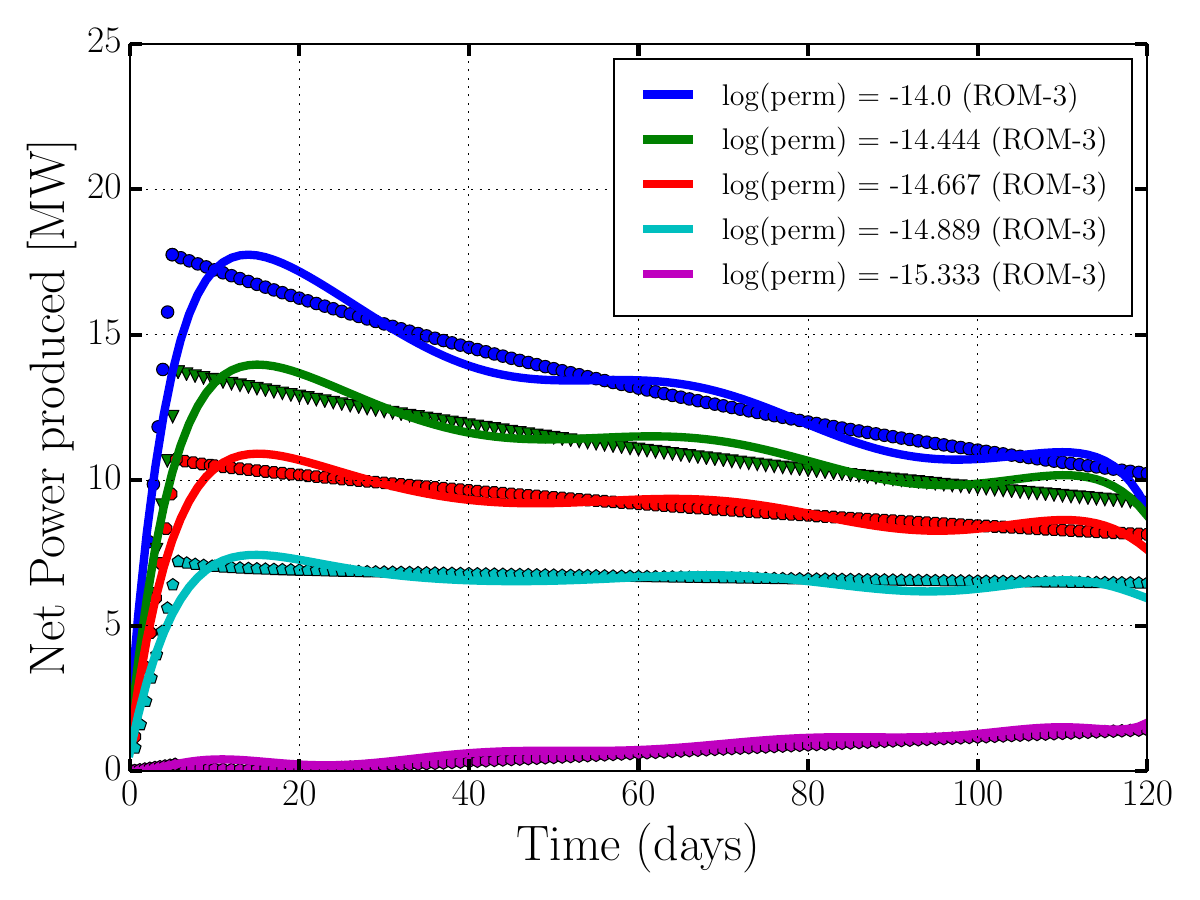}}
  \subfigure[\textbf{Validation:} PFLOTRAN 
    simulations and ROM-3]
    {\includegraphics[scale=0.4]
    {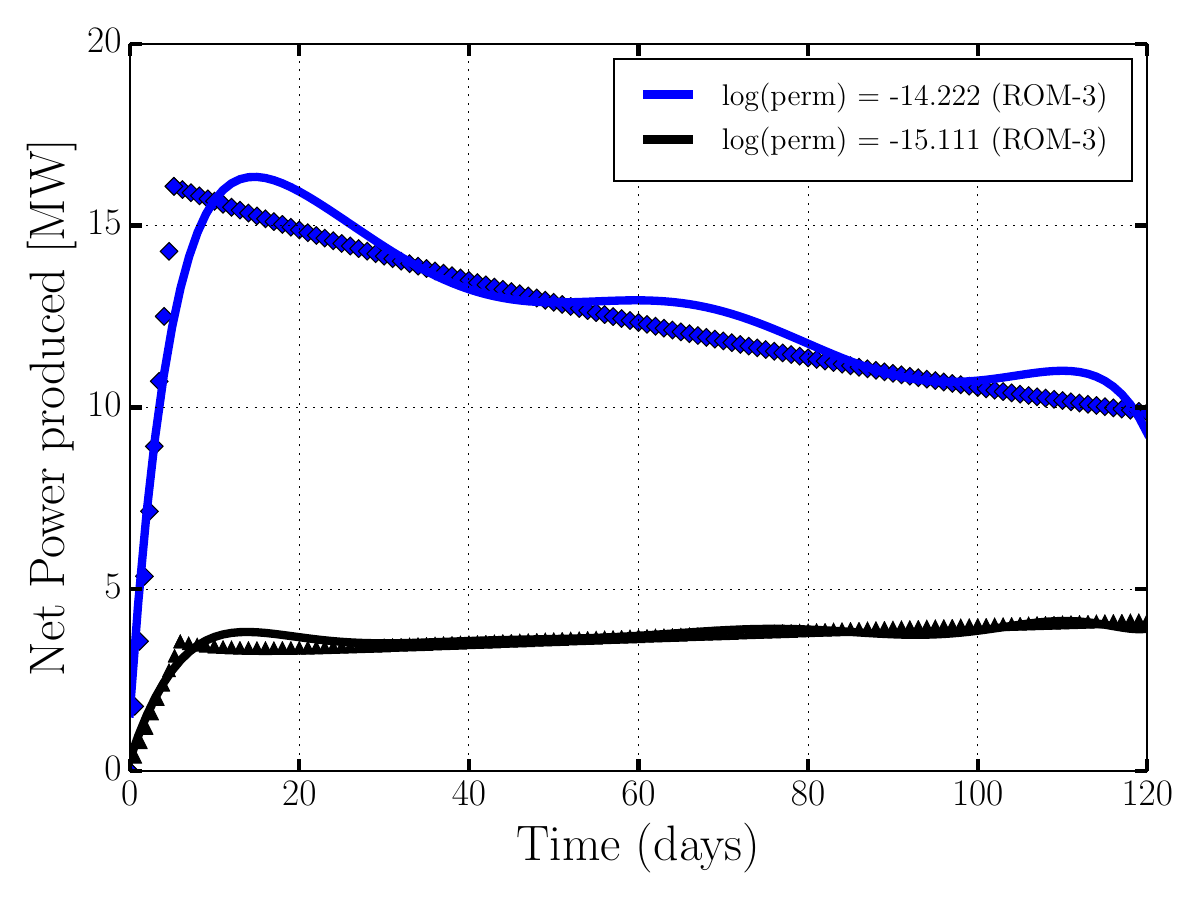}}
  \subfigure[\textbf{Prediction:} Comparison of LTFT power 
    data and ROM-3]
    {\includegraphics[scale=0.4]
    {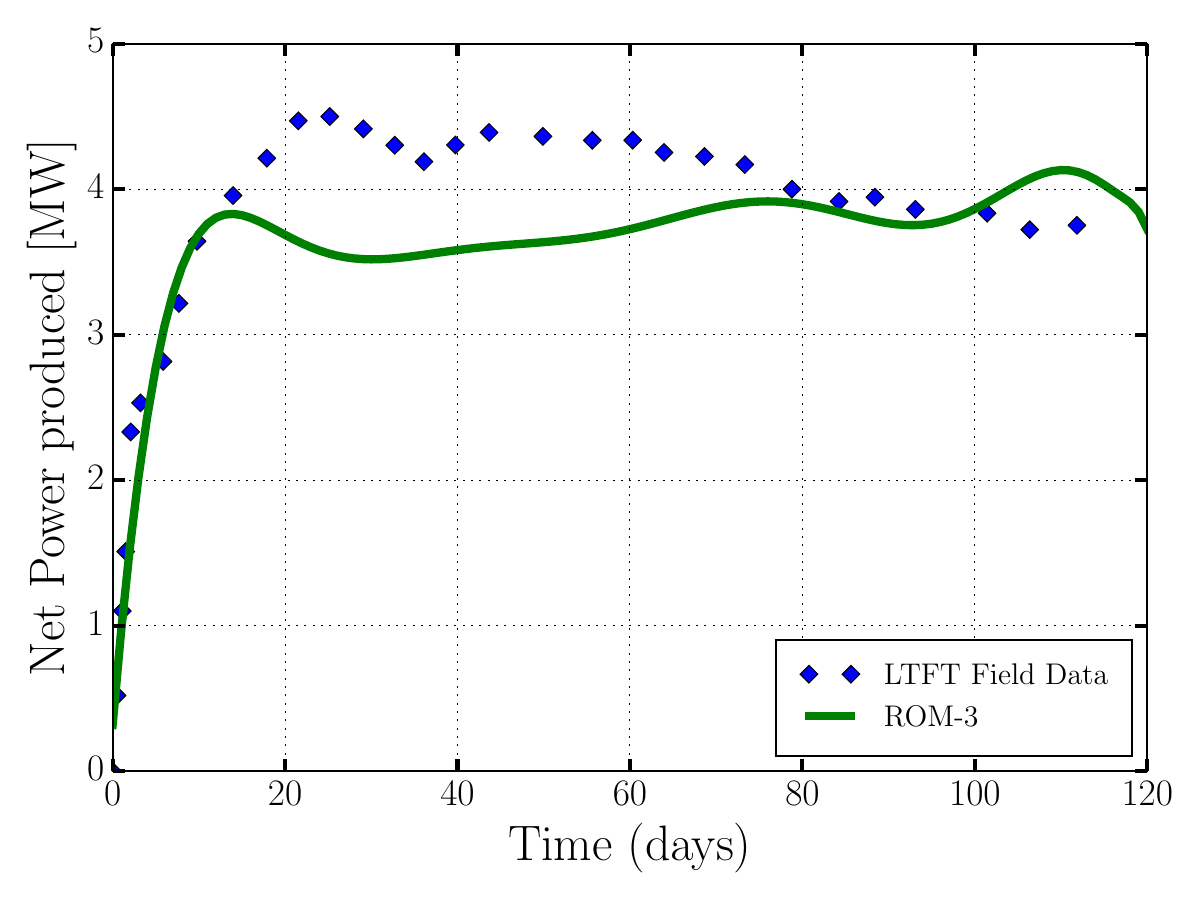}}
  \caption{\textsf{Thermal power output ROM-3:}~The top 
    left figure compares the ROM-3 output with PFLOTRAN numerical 
    simulations over a training set. Numerical simulation data is shown with markers and 
    ROMs are shown with solids of same color.
    For training, the $\mathrm{R}^2$-values 
    are equal to 0.917, 0.896, 0.9, 0.889, and 0.813 for log(perm) values 
    of -14.0, -14.444, -14.667, -14.889, and -15.333. The top right figure 
    is the validation of the proposed ROM-3 with the PFLOTRAN simulations. 
    For validation, the $\mathrm{R}^2$-values are equal to 0.892 and 0.893 
    for log(perm) values of -14.222 and -15.111. The entire data set consists 
    of 7 numerical simulations out of which 5 simulations are used for training 
    and 2 simulations are used for validation of the developed ROM-3. The bottom 
    figure compares the predictions of ROM-3 with LTFT field-scale power output 
    data set of Phase II experiment. The fracture zone permeability 
    used is the calibrated value for constant injection flow rate. The 
    $\mathrm{R}^2$-value for this prediction is equal to 0.824. In essence, 
    qualitatively and quantitatively, ROM-3 is able to describe PFLOTRAN 
    simulations at all given ranges of permeabilities. Even though ROM-3 
    is not exactly a close match to the LTFT Phase II data qualitatively, 
    it is a much better model compared to ROM-1 due to incorporation of 
    higher-order polynomials.
  \label{Fig:PFLOTRAN_LTFTdata_ROM3}}
\end{figure}

Motivated by these two ROMs, ROM-3 is constructed. The philosophy of ROM-3 is to 
use only polynomials. The number of terms is determined by the coefficient values. 
The maximum possible order for the polynomial chosen is 10. This is because as the 
order of polynomial increases the values of the coefficient are close to machine 
precision (close to zero). Figure \ref{Fig:PFLOTRAN_LTFTdata_ROM3} 
shows the training, validation using numerical simulations of ROM-3 along with 
its predictions. The predictions are compared with field power data. From 
this figure (8a/9a), it is apparent that ROM-3 more accurately reproduces the trend 
in the behavior of thermal power output than ROM-2. For numerical simulations, as 
the permeability increases the deviations in the output values of ROM-3 and PFLOTRAN 
numerical simulations are not very large as compared to ROM-1 and ROM-2. In the case 
of the LTFT experiment, ROM-3 is able to accurately describe the increase in the thermal 
power output in initial stages. After time $t = 20$ days, when compared to the performance 
of ROM-2, ROM-3 is not exactly a close match to the LTFT data quantitatively. However, 
qualitatively, ROM-3 is a much better model compared to ROM-1 due to the incorporation 
of higher-order polynomials (see Tables \ref{Tab:ROMs_Summary_1} 
and \ref{Tab:ROMs_Summary_2} for more details). In short, 
ROM-3 neither overfits nor underfits the LTFT experiment data. Hence, ROM-3 is a better 
model compared to ROM-2 and ROM-1 and provides a middle ground for model parsimony. The 
model can be improved by incorporating other input terms such as well factor, bottom hole 
pressure, and injection mass flow rates. This is beyond the scope of the current paper 
and will be considered in our future work. Table \ref{Tab:ROMs_Summary_1} summarizes the ROMs 
and their $\mathrm{R}^2$-values for training, validation, and prediction of numerical 
simulations and LTFT field-data.

To conclude the discussion, the following can be inferred based on Figures 
\ref{Fig:PFLOTRAN_LTFTdata_ROM1}--\ref{Fig:PFLOTRAN_LTFTdata_ROM3} and 
equations \eqref{Eqn:ROM_1}--\eqref{Eqn:ROM3_Coeff3}:
\begin{itemize}
  \item [$\blacktriangleright$] In reproducing PFLOTRAN numerical simulations, 
    for low values of permeability, ROM-1 outperforms ROM-2 and ROM-3. At 
    higher values of permeability, ROM-2 outperforms ROM-1 and ROM-3. In 
    predicting LTFT data, ROM-2 outperforms ROM-3 and ROM-1. However, ROM-3 
    is able to describe the initial trend in the field-data and other qualitative 
    aspects (such as the rise in power production after $t = 10$ days and 
    decline after $t = 30$ days).
  \item [$\blacktriangleright$] At first glance, it may seem that ROM-2 
    may be the best model in reproducing LTFT data. However, from Figure 
    \ref{Fig:PFLOTRAN_LTFTdata_ROM2}, it is evident that ROM-2 considerably 
    deviates from the PFLOTRAN simulations at low values of permeability. 
    Typically, for EGS applications, higher effective fracture zone permeabilities 
    are desired \cite{Brown_etal}. This is because in many practical scenarios 
    (assuming that reservoir matrix permeability to be very low), higher fracture 
    zone permeabilities can be correlated to (possibly) well connected and dispersed 
    discrete fracture networks \cite{Adler_etal}. This means that the fluid sweeps 
    through a larger fractured volume (as compared to lower fracture zone permeabilities) 
    resulting in higher power production at producing wells. To model such scenarios, 
    we believe ROM-2 and ROM-3 are better models than ROM-1. 
\end{itemize}
%

\section{CONCLUDING REMARKS}
\label{Sec:S4_Geothermal_Conclusions}
In this paper, we have presented various reduced-order models to 
describe different aspects of the numerical simulations and LTFT 
field-scale power output data of Phase II Fenton hill geothermal 
reservoir. First, we described the governing equations for fluid 
flow in the fractured reservoir and corresponding thermal drawdown. 
Second, we have presented a physics-based conceptual model for an 
EGS reservoir. The conceptual model is an approximation of a more 
complex system, which is used to understand the essential features 
of the systems and make it amenable for numerical simulations. 
Field-scale data sets, which are extracted from the documents 
provided by the geothermal code comparison project, are used 
to estimate the parameters of the EGS system under consideration. 
These data sets include pressures, backpressures, mass flow rates, 
and temperatures at both injection and production sites. Third, 
sensitivity analysis is performed on these inputs to identify 
and rank the key parameters:~fracture zone permeability, 
well factor, injection flow rate and bottom hole pressure. Based on 
this analysis, fracture zone permeability was found to be most important and so we 
used only permeability and time as the parameters for ROM construction. 
Finally, the ROMs are developed using the numerical simulations obtained 
based on equally spaced parametric values.

The ROMs are built (trained and validated) using 
simulation data from the numerical model (PFLOTRAN) as shown 
in Figures \ref{Fig:PFLOTRAN_LTFTdata_ROM1}, \ref{Fig:PFLOTRAN_LTFTdata_ROM2}, 
and \ref{Fig:PFLOTRAN_LTFTdata_ROM3}. Then, these ROMs are 
used to predict the behavior of an EGS system, specifically, 
Fenton Hill system. The input to these ROMs are permeability 
and time. From the calibration in Figure 4, we first obtain 
the permeability of the Fenton Hill EGS system. This permeability 
value along with time are then used in ROM predictions, and 
compared with the observation data to evaluate the performance 
of the ROMs.

We evaluated three different ROMs with different levels of model 
parsimony, each describing key and essential features of the LTFT 
power output data. The first ROM is a simple model and is able to 
accurately describe the power output at low fracture zone permeabilities, 
and is relatively parsimonious. The second ROM is a more complex model 
than ROM-1. However, ROM-2 shares some of the limitations of ROM-1. 
In general, ROM-2 reproduces the numerical simulations better than 
ROM-1 at higher permeabilities. The interesting part of ROM-2 is 
that the power output predicted closely matches the LTFT experiment, 
qualitatively and quantitatively. The third ROM is constructed by 
taking the best aspects of ROM-1 and ROM-2, and provides a middle 
ground for model parsimony. ROM-3 is able to quantitatively and 
qualitatively describe the trend in the power output at different 
time levels for both PFLOTRAN numerical simulations and LTFT power 
output data.

From these ROM development workflows and sensitivity analyses, it 
is evident that this study has demonstrated that simple reduced-order 
models are able to capture various complex features in the system. 
This work provides confidence in developing simple and efficient 
transient reduced-order models for geothermal field use. For EGS 
applications, higher fracture zone permeability is desired \cite{Brown_etal}. 
This is because at higher permeabilities power output is higher 
as the fluid sweeps through a larger fracture zone volume. For 
such scenarios (at higher permeabilities in predicting the thermal 
power production), ROM-2 and ROM-3 outperform ROM-1. We think ROM-2 
and ROM-3 show promise for EGS studies. 

\section*{APPENDIX}
\label{Sec:S5_Geothermal_Appendix}
The coefficients for ROM-1 are given as follows:
\begin{subequations}
  \begin{align}
    \label{Eqn:ROM1_Coeff0}
    a_0(t) &= -2.689 \times 10^3 - 9.951 \times 10^2 t + 
    28.37 t^2 - 3.013 \times 10^{-1} t^3 + 1.095 \times 
    10^{-3} t^4\\
    \label{Eqn:ROM1_Coeff1}
    a_1(t) &= 5.517 \times 10^2 + 2.01 \times 10^2 t - 
    5.751 t^2 - 6.111 \times 10^{-2} t^3 - 2.222 \times 
    10^{-4} t^4 \\
    \label{Eqn:ROM1_Coeff2}
    a_2(t) &= -3.718 \times 10^1 - 1.347 \times 10^1 t + 
    3.867 \times 10^{-1} t^2 - 4.112 \times 10^{-3} t^3 
    + 1.495 \times 10^{-5} t^4 \\
    \label{Eqn:ROM1_Coeff3}
    a_3(t) &= 8.241 \times 10^3 + 2.998 \times 10^{-1} t - 
    8.634 \times 10^{-3} t^2 + 9.184 \times 10^{-5} t^3 
    - 3.339 \times 10^{-7} t^4
  \end{align}
\end{subequations}

For ROM-2, the coefficients $b_i$ are functions of time (days), 
which are given as follows:
\begin{subequations}
  \begin{align}
    \label{Eqn:ROM2_Coeff0}
    b_0(t) &= 2.318 \times 10^3 - 2.466 \times 10^3 t + 
    1.338 \times 10^2 t^2 - 3.413 t^3 + 4.612 \times 
    10^{-2} t^4 \nonumber \\
    &- 3.393 \times 10^{-4} t^5 + 1.376 \times 10^{-6} 
    t^6 - 3.538 \times 10^{-9} t^7 + 6.869 \times 10^{-12} 
    t^8 \nonumber \\
    &- 2.307 \times 10^3 \times (0.1)^t + 1.342 \sin(t) \\
    \label{Eqn:ROM2_Coeff1}
    b_1(t) &= -4.623 \times 10^2 + 4.992 \times 10^2 t - 
    2.713 \times 10^1 t^2 + 6.918 \times 10^{-1} t^3 -
    9.339 \times 10^{-3} t^4 \nonumber \\
    &+ 6.856 \times 10^{-5} t^5 - 2.766 \times 10^{-7} 
    t^6 + 7.056 \times 10^{-10} t^7 - 1.369 \times 10^{-12} 
    t^8 \nonumber \\
    &- 2.307 \times 10^3 \times (0.1)^t -2.676 \sin(t) \\
    \label{Eqn:ROM2_Coeff2}
    b_2(t) &= 3.103 \times 10^1 - 3.353 \times 10^1 t + 
    1.825 t^2 - 4.652 \times 10^{-2} t^3 + 6.28 \times 
    10^{-4} t^6 - 4.608 \times 10^{-6} t^5 
    \nonumber \\
    &+ 1.858 \times 10^{-8} t^6 - 4.734 \times 10^{-11} 
    t^7 + 9.191 \times 10^{-14} t^8 - 3.087 \times 10^1 
    \times (0.1)^t \nonumber \\
    &+ 1.796 \times 10^{-2} \sin(t) \\
    \label{Eqn:ROM2_Coeff3}
    b_3(t) &= -6.997 \times 10^{-1} + 7.477 \times 10^{-1} t - 
    4.073 \times 10^{-2} t^2 + 1.038 \times 10^{-3} t^3 -
    1.402 \times 10^{-5} t^4 \nonumber \\
    &+ 1.031 \times 10^{-7} t^5 - 4.168 \times 10^{-10} t^6 
    -1.068 \times 10^{-12} t^7 - 2.073 \times 10^{-15} t^8 
    \nonumber \\
    &+ 6.964 \times 10^{-1} \times (0.1)^t - 4.048 \times 
    10^{-4} \sin(t)
  \end{align}
\end{subequations}
The parameters $t_j$ (days) are chosen in such 
a way that the ROM outputs be close to that of the field-scale 
LTFT thermal power output at the $i$-th time-snapshots/time-levels. 
These time values are given as follows:
\begin{subequations}
  \begin{align}
    \label{Eqn:ROM2_Param1}
    &t_1 = 10, t_2 = 15, t_3 = 17.5, t_4 = 19, t_5 = 20, t_6 = 21, 
    t_7 = 22.5, t_8 = 25, t_9 = 30, t_{10} = 32.5 \\
    \label{Eqn:ROM2_Param2}
    &t_{11} = 35, t_{12} = 37.5, t_{13} = 50, t_{14} = 52.5, t_{15} 
    = 55, t_{16} = 60, t_{17} = 62.5, t_{18} = 65, t_{19} = 70, \\
    \label{Eqn:ROM2_Param3}
    &t_{20} = 75, t_{21} = 80, t_{22} = 85, t_{23} = 87.5, t_{24} = 90, t_{25} 
    = 95, t_{26} = 100, t_{27} = 105, \\
    \label{Eqn:ROM2_Param4}
    &t_{28} = 110, t_{29} = 112.5
  \end{align}
\end{subequations}
The coefficients $m_j, n_j, r_j$ are given as follows:
\begin{subequations}
  \begin{align}
    \label{Eqn:ROM2_Param3}
    &m_1 = 0.1, m_2 = 0.085, m_3 = 0.05, m_4 = -0.0125, 
    m_5 = 0.125, m_6 = 0.1, m_7 = -0.0125 \\
    \label{Eqn:ROM2_Param4}
    &m_8 = 0.085, m_9 = -0.075, m_{10} = -0.085, m_{11} 
    = 0.085, m_{12} = 0.125, m_{13} = -0.085, \\
    \label{Eqn:ROM2_Param5}
    &m_{14} = -0.01, m_{15} = -0.05, m_{16} = -0.075, 
    m_{17} = -0.075, m_{18} = -0.025, m_{19} = -0.015, \\
    \label{Eqn:ROM2_Param6}
    &m_{20} = -0.15, m_{21} = -0.05, m_{22} = -0.05, 
    m_{23} = -0.05, m_{24} = -0.15, m_{25} = -0.1, \\
    \label{Eqn:ROM2_Param7}
    &m_{26} = -0.085, m_{27} = -0.175, m_{28} = -0.05, 
    m_{29} = 0.05, \\
    \label{Eqn:ROM2_Param7}
    &n_1 = n_2 = n_3 = 100, \; n_4 \; \mathrm{to} \; n_{29} = 1000 \\
    \label{Eqn:ROM2_Param8}
    &r_1 = r_2 = r_3 = 1, r_4 = 0.5, r_5 = 0.25, r_6 = 0.5, 
    r_7 = 0.1, r_8 = 0.25, r_9 = 0.75, r_{10} = 0.02 \\
    \label{Eqn:ROM2_Param9}
    &r_{11} = 0.01, r_{12} = 0.01, r_{13} = 0.25, r_{14} = 
    1.5, r_{15} = 1.75, r_{16} = r_{17} = r_{18} = r_{19} = 
    r_{20} = 0.01 \\
    \label{Eqn:ROM2_Param10}
    &r_{21} = r_{22} = r_{23} = 0.01, r_{24} = 0.025, r_{25} 
    = 0.075, r_{26} = 0.01, r_{27} = 0.15, \\
    \label{Eqn:ROM2_Param11}
    &r_{28} = 0.01, r_{29} = 0.01
  \end{align}
\end{subequations}

For ROM-3, the coefficients $c_i$ are functions of time (days), which 
are given as follows:
\begin{subequations}
  \begin{align}
    \label{Eqn:ROM3_Coeff0}
    c_0(t) &= 7.913 \times 10^2 - 1.747 \times 10^3 t + 
    2.089 \times 10^1 t^2 + 5.173 t^3 - 3.234 \times 
    10^{-1} t^4 \nonumber \\
    &+ 9.397 \times 10^{-3} t^5 - 1.613 \times 10^{-4} t^6 
    + 1.727 \times 10^{-6} t^7 - 1.134 \times 10^{-8} t^8 
    \nonumber \\
    &+ 4.183 \times 10^{-11} t^9 - 6.627 \times 10^{-14} t^{10} \\
    \label{Eqn:ROM3_Coeff1}
    c_1(t) &= -1.578 \times 10^2 + 3.557 \times 10^2 t - 
    4.613 t^2 - 1.093 t^3 + 6.432 \times 10^{-2} t^4  
    \nonumber \\
    &-1.872 \times 10^{-3} t^5 + 3.215 \times 10^{-5} t^6 
    - 3.442 \times 10^{-7} t^7 + 2.261 \times 10^{-9} t^8 
    \nonumber \\
    &- 8.337 \times 10^{-12}t^9 + 1.321 \times 10^{-14} t^{10} \\
    \label{Eqn:ROM3_Coeff2}
    c_2(t) &= 1.059 \times 10^1 - 2.391 \times 10^1 t - 
    3.136 \times 10^{-1} t^2 + 6.835 \times 10^{-2} t^3 
    - 4.316 \times 10^{-3} t^4 \nonumber \\
    &+ 1.256 \times 10^{-4} t^5 - 2.158 \times 10^{-6} t^6 
    + 2.311 \times 10^{-8} t^7 - 1.517 \times 10^{-10} t^8 
    \nonumber \\
    &+ 5.597 \times 10^{-13}t^9 - 8.868 \times 10^{-16} t^{10} \\
    \label{Eqn:ROM3_Coeff3}
    c_3(t) &= -2.388 \times 10^{-1} + 5.305 \times 10^{-1} t - 
    6.637 \times 10^{-3} t^2 - 1.553 \times 10^{-3} t^3 
    + 9.754 \times 10^{-5} t^4 \nonumber \\
    &- 2.837 \times 10^{-6} t^5 + 4.871 \times 10^{-8} t^6 
    - 5.215 \times 10^{-10} t^7 + 3.425 \times 10^{-12} t^8 
    \nonumber \\
    &-1.263 \times 10^{-14} t^9 + 2.001 \times 10^{-17} t^{10}
  \end{align}
\end{subequations}

\section*{ACKNOWLEDGMENTS}
The authors thank U.S. Department of Energy (DOE) - Geothermal 
Technologies Program Office  for support through project DE-AC52-06NA25396.
MKM and SK also thank LANL Laboratory Directed
Research and Development for the support through
Early Career Project 20150693ECR.
MKM thanks Don Brown and 
Sharad Kelkar for many useful and knowledgeable discussions. 
The authors also thank two anonymous reviewers for their feedback
which helped improve the manuscript.

\bibliographystyle{unsrt}
\bibliography{Master_References/Master_References,Master_References/Books}

\end{document}